\documentclass[12pt]{article}
\usepackage{ametsoc}
\newcommand{\Sig}{\mbox{\boldmath \(\Sigma\)}}
  \newcommand{\PN}{\textrm{PN}}
  \newcommand{\DO}{do}
  \newcommand{\PS}{\textrm{PS}}
  \newcommand{\PNS}{\textrm{PNS}}

\newcommand{\Proba}{\mathbb P}
\newcommand{\Esp}{\mathbb E}
\newcommand{\Var}{\textrm{V}}
\newcommand{\x}{\boldsymbol{x}}
\newcommand{\vv}{\textrm{v}}
\newcommand{\Om}{\boldsymbol{\Omega}}
\newcommand{\E}{\mathbf{E}}
\newcommand{\D}{\mathbf{\Delta}}

\newcommand{\Sn}{\widehat{\mathbf{\Omega}}}

\newcommand{\myabstract}{Multiple changes in Earth's climate system have been observed over the past decades. Determining how likely each of these changes are to have been caused by human influence, is important for decision making on mitigation and adaptation policy. Here we describe an approach for deriving the probability that anthropogenic forcings have caused a given observed change. The proposed approach is anchored into causal counterfactual theory \citep{Pearl09} which has been introduced recently, and was in fact partly used already, in the context of extreme weather event attribution (EA). We argue that these concepts are also relevant, and can be straightforwardly extended to, the context of detection and attribution of long term trends associated to climate change (D\&A). For this purpose, and in agreement with the principle of \textit{fingerprinting} applied in the conventional D\&A framework, a trajectory of change is converted into an event occurrence defined by maximizing the causal evidence associated to the forcing under scrutiny. Other key assumptions used in the conventional D\&A framework, in particular those related to numerical models error, can also be adapted conveniently to this approach. Our proposal thus allows to bridge the conventional framework with the standard causal theory, in an attempt to improve the quantification of causal probabilities. An illustration suggests that our approach is prone to yield a significantly higher estimate of the probability that anthropogenic forcings have caused the observed temperature change, thus supporting more assertive causal claims.}
\begin{document}
%
%%%%%%%%%%%%%%%%%%%%%%%%%%%%%%%%%%%%%%%%%%%%%%%%%%%%%%%%%%%%%%%%%%%%%
% TITLE
%
% Enter your TITLE here
%%%%%%%%%%%%%%%%%%%%%%%%%%%%%%%%%%%%%%%%%%%%%%%%%%%%%%%%%%%%%%%%%%%%%
\title{\textbf{\large{Probabilities of causation of climate changes}}}
%
% Author names, with corresponding author information. 
% [Update and move the \thanks{...} block as appropriate.]
%
\author{\textsc{Alexis Hannart}
				\thanks{\textit{Corresponding author address:} 
				Alexis Hannart, OURANOS, 550 rue Sherbrooke Ouest, Montreal, QC, Canada. 
				\newline{E-mail: hannart.alexis@ouranos.ca}}\\%\quad\textsc{Jack Crielson, and Sarah Cooley}\\
\textit{\footnotesize{OURANOS, Montreal, Canada}}\\
\textit{\footnotesize{IFAECI, CNRS/CONICET/UBA, Buenos Aires, Argentina}}
\and 
\centerline{\textsc{Philippe Naveau}}\\% Add additional authors, different insitution
\centerline{\textit{\footnotesize{LSCE, CNRS/CEA, Gif-sur-Yvette, France}}}
}
%
% Formatting done here...Authors should skip over this.  See above for abstract.
\ifthenelse{\boolean{dc}}
{
\twocolumn[
\begin{@twocolumnfalse}
\amstitle

% Start Abstract (Enter your Abstract above.  Do not enter any text here)
\begin{center}
\begin{minipage}{13.0cm}
\begin{abstract}
	\myabstract
	\newline
	\begin{center}
		\rule{38mm}{0.2mm}
	\end{center}
\end{abstract}
\end{minipage}
\end{center}
\end{@twocolumnfalse}
]
}
{
\amstitle
\begin{abstract}
\myabstract
\end{abstract}
\newpage
}
%%%%%%%%%%%%%%%%%%%%%%%%%%%%%%%%%%%%%%%%%%%%%%%%%%%%%%%%%%%%%%%%%%%%%
% MAIN BODY OF PAPER
%%%%%%%%%%%%%%%%%%%%%%%%%%%%%%%%%%%%%%%%%%%%%%%%%%%%%%%%%%%%%%%%%%%%%

\section{Introduction}
Investigating causal links between climate forcings and the observed climate evolution over the instrumental era represents a significant part of the research effort on climate. Studies addressing these aspects in the context of climate change have been providing over the past decades, an ever increasing level of causal evidence that is important for decision-makers in international discussions on mitigation policy. In particular, these studies have produced far-reaching causal claims; for instance the latest IPCC report \citep{AR5}, AR5 hereafter, stated that ``\textit{It is extremely likely that human influence has been the dominant cause of the observed warming since the mid-20$^{\textrm{th}}$ century}." An important part of this causal claim, as well as many related others, regards the associated level of uncertainty. More precisely, the expression ``\textit{extremely likely}" in the latter quote has been formally defined by the IPCC \citep{Mas10} to correspond to a probability of $95\%$. The above quote hence implicitly means that the probability that the observed warming since the mid-20th century was not predominantly caused by human influence but by natural factors, is roughly $1:20$. Based on the current state of knowledge, it means that it is not yet possible to fully rule out that natural factors were the main causes of the observed global warming. %In an hypothetical lawsuit on global warming, to follow a legal analogy, mankind could therefore arguably not be declared responsible ``\textit{beyond reasonable doubt}" based on the available evidence --- at least according to the above claim and by using a fairly conservative threshold for ``reasonable doubt''. 
This probability of $1:20$, as well as all the probabilities associated to the numerous causal claims that can be found in the past and present climate literature and are summarized in AR5, are critical quantities that are prone to affect the way in which climate change is apprehended by citizens and decision makers, and thereby to affect decisions on the matter. It is thus of interest to examine the method followed to derive them and, potentially, to improve it.

Aforementioned studies buttressing the above claim usually rely on a conventional attribution framework in which ``\textit{causal attribution of anthropogenic climate change}'' is understood to mean ``\textit{demonstration that a detected change is consistent with the estimated responses to anthropogenic and natural forcings combined, but not consistent with alternative, physically plausible explanations that exclude important elements of anthropogenic forcings}'' \citep{Guidance}. While this definition has proved to be very useful and relevant, it offers a description of causality which is arguably overly qualitative for the purpose of deriving a probability. In particular, it comes short of a mathematical definition of the word ``\textit{cause}'' and incidentally, of the ``\textit{probability to have caused}'' that we in fact wish to quantify. Hence, beyond these general guidance principles, the actual derivation of these probabilities is left to some extent to the interpretation of the practitioner. In practice, causal attribution has usually been performed by using a class of linear regression models \citep{HZ11}: 
\begin{equation}\label{regress}
y=\sum_{f=1}^p \beta_f x_f+\varepsilon
\end{equation}
where the observed climate change $y$ is regarded as a linear combination of $p$ externally forced response patterns $x_f$ with $f=1,..., p$ referred to as fingerprints, and where $\varepsilon$ represent of internal climate variability and observational error (all variables are vectors of dimension $n$). The regression coefficient $\beta_f$ accounts for possible error in climate models in simulating the amplitude of the pattern of response to forcing $f$. After inference and uncertainty analysis, the value of each coefficient $\beta_f$ and the magnitude of the confidence intervals determine whether or not the observed response is attributable to the associated forcing. The desired probability of causation, i.e. the probability that the forcing of interest $f$ has caused the observed change $y$ is denoted hereafter $\Proba(f\rightarrow y)$. It has often been equated to the probability that the corresponding linear regression coefficient is positive\footnote{The notation $\Proba(\beta_f>0)$ corresponds to the confidence level associated to the confidence interval $[0,+\infty[$ under a frequentist approach, and to the posterior probability that $\beta_f$ is positive under a Bayesian one. %wherein $\beta_f$ is a random variable, 
%and can be defined as the % formulation wherein $\beta_f$ is a constant.
}:
\begin{equation}\label{defconv}
%\begin{align}
\Proba(f\rightarrow y) = \Proba(\beta_f>0)
%& \Proba(f\rightarrow y) = \Proba(a_f<1)
%\end{align}
\end{equation}
A shortcoming of the conventional framework summarized in Equations (\ref{regress}) and (\ref{defconv}) above, is that a linear regression coefficient does not have any causal meaning from a formal standpoint. %Therefore, the method used to derive the probabilities of causation of climate change can be described, in our view, as heuristic insofar as it does not formally quantify a probability of causation. 
%It is perhaps worth noting upfront that, in respect to this shortcoming, climate science is not an exception: scientists in all fields have indeed long had difficulties in defining precisely causality \citep{Pearl09}. Furthermore, 
As acknowledged by \cite{Pearl09}, turning an intrinsically deterministic notion such as causality into a probabilistic one, is a difficult general problem which has also long been a matter of debate \citep{Sim51,Sup70,Mel95}. %Arguably, heuristic definitions of causality have therefore long been commonplace in many scientific fields.
Nevertheless, the current approach can be theoretically improved in the context of climate change where the values of the probabilities of causation have such important implications. 

Our proposal to tackle this objective is anchored into a coherent theoretical corpus of definitions, concepts and methods of general applicability which has emerged over the past three decades to address the issue of evidencing causal relationships empirically \citep{Pearl09}. This general framework is increasingly used in diverse fields (e.g. in epidemiology, economics, social science) in which investigating causal links based on observations is a central matter. Recently, it has been introduced in climate science for the specific purpose of attributing weather and climate-related extreme events \citep{HPONG15}, which we refer to as `extreme events' hereafter. The latter article gave a brief overview of causal theory and articulated it with the conventional framework used for the attribution of extreme events, which is also an important topic in climate attribution. In particular, \cite{HPONG15} showed that the key quantity referred to as the fraction of attributable risk (FAR) \citep{All03,Stone05} which buttresses most extreme event attribution (EA) studies, can be directly interpreted within causal theory.
%he quantity $1-\overline{p}/p$ in Equation (\ref{prob2}), which has indeed been introduced in this field one decade ago as the fraction of attributable risk (FAR) by \cite{All03,Stone05}; 

However, \cite{HPONG15} did not address how to extend and adapt this theory in the context of the attribution of climate changes occurring on longer timescales. Yet, a significant advantage of the definitions of causal theory is precisely that they are relevant no matter the temporal and spatial scale. %They can be identically applied to the causal attribution of observed changes in the climate system occurring on the long term, and to the attribution of a single weather or climate-related event. 
For instance, from the perspective of a paleoclimatologist studying Earth's climate over the past few hundred millions of years, global warming over the past hundred and fifty years can be considered as a climate event. As a matter of fact, the word ``event'' is used in paleoclimatology to refer to ``rapid'' changes in the climate system, but ones that may yet last centuries to millennia. %Hence, defining the distinction between attributing a climate event and attributing a climate trend is thus arguably a judgement call. 
Where to draw the line %between the short lead times associated to events and the long ones associated to trends 
is thus arbitrary: one person's long term trend is another person's short term event. It should therefore be possible to tackle causal attribution within a unified methodological framework based on shared concepts and definitions of causality. Doing so would allow to bridge the methodological gap that exists between EA and trend attribution at a fundamental level, thereby covering the full scope of climate attribution studies. Such a unification would present in our view several advantages: enhancing methodological research synergies between D\&A topics, improving the shared interpretability of results, and streamlining the communication of causal claims --- in particular when it comes to the quantification of uncertainty, i.e. of the probability that a given forcing has caused a given observed phenomenon.

Here, we adapt some formal definitions of causality and probability of causation to the context of climate change attribution. Then, we detail technical implementation under standard assumptions used in D\&A. The method is finally illustrated on the warming observed over the 20$^{\textrm{th}}$ century.

\section{Causal counterfactual theory} 

While an overview of causal theory cannot be repeated here, it is necessary for clarity and self-containedness to highlight its key ideas and most relevant concepts for the present discussion.

Let us first recall the so-called ``counterfactual'' definition of causality by quoting the 18th century Scottish philosopher David Hume: ``\textit{We may define a cause to be an object followed by another, where, if the first object had not been, the second never had existed.}''  In other words, an event $E$ ($E$ stands for effect) is caused by an event $C$ ($C$ stands for cause) if and only if $E$ would not occur were it not for $C$. Note that the word \textit{event} is used here in its general, mathematical sense of \textit{subset} of a sample space $\Omega$. According to this definition, evidencing causality requires a counterfactual approach by which one inquires whether or not the event $E$ would have occurred in a hypothetical world, termed counterfactual, in which the event $C$ would not have occurred. The fundamental approach of causality which is implied by this definition is still entirely relevant in the standard causal theory. It may also arguably be connected to the guidance principles of the conventional climate change attribution framework and to the optimal fingerprinting models, in a qualitative manner. %It also explicitly buttresses the conventional framework used for event attribution \citep{Allen04}. The conventional framework used for 
The main virtue of the standard causality theory of Pearl consists in our view in formalizing precisely the above qualitative definition, thus allowing for sound quantitative developments. A prominent feature of this theory consists in first recognizing that causation corresponds to rather different situations and that three distinct facets of causality should be distinguished: (i) necessary causation, where the occurrence of $E$ requires that of $C$ but may also require other factors; (ii) sufficient causation, where the occurrence of $C$ drives that of $E$ but may not be required for $E$ to occur; (iii) necessary and sufficient causation, where (i) and (ii) both hold. The fundamental distinction between these three facets can be visualized by using the simple illustration shown in Figure 1. 

While the counterfactual definition as well as the three facets of causality described above may be understood at first in a fully deterministic sense, perhaps the main strength of Pearl's formalization is to propose an extension of these definitions under a probabilistic setting. The probabilities of causation are thereby defined as follow:
\begin{subequations}\label{prob}
\begin{align}
& \PS(C\rightarrow E) = \Proba(E\mid \DO(C),\overline{C},\overline{E})\,, \\
& \PN(C\rightarrow E) = \Proba(\overline{E}\mid \DO(\overline{C}),C,E)\,, \\
& \PNS(C\rightarrow E) = \Proba(E\mid\DO(C),\overline{E}\mid \DO(\overline{C}))\,.
\end{align}
\end{subequations}
where $\overline{C}$ and $\overline{E}$ are the complementaries of $C$ and $E$, and where the notation $\DO(.)$ means that an \textit{intervention} is applied to the system under causal investigation. For instance $\PS$, the \textit{probability of sufficient causation}, reads from the above: the probability that $E$ occurs when $C$ is interventionally forced to occur, conditional on the fact that neither $C$ nor $E$ were occurring in the first place. Conversely $\PN$, the \textit{probability of necessary causation}, is defined as the probability that $E$ would not occur when $C$ is interventionally forced to not occur, conditional on the fact that both $C$ and $E$ were occurring in the first place. %Note that both $\PN$ and $\PS$ are counterfactual probabilities in the sense that they involve interventions to explore hypothetical situations that did not occur. 
While we omit here the formal definition of the intervention $\DO(.)$ for brevity, the latter can be understood merely as experimentation: applying these definitions thus requires the ability to experiment. Real experimentation, whether \textit{in situ} or \textit{in vivo}, is often accessible in many fields but it is not in climate research for obvious reasons. In this case, one can thus only rely on numerical  \textit{in silico} experimentation:  the implications of this constraint are discussed further. 

While the probabilities of causation are not easily computable in general, their expression fortunately becomes quite simple under assumptions that are reasonable in the case of external forcings (i.e. exogeneity and monotonicity):
\begin{subequations}\label{prob2}
\begin{align}
& \textrm{PN}(C\rightarrow E) =\max \left(1 - \overline{p}/p, 0\right), \\%\label{proba1} 
& {\textrm{PS}}(C\rightarrow E) =\max \left(1- (1-p)/(1-\overline{p}), 0\right)\,, \\
& \textrm{PNS}(C\rightarrow E) = \max \left(p-\overline{p}, 0\right)\,.
\end{align}
\end{subequations}
where $p=\Proba(E\mid \DO(C))$ is the so-called \textit{factual} probability of the event $E$ in the real world where $C$ did occur and $\overline{p}=\Proba(E\mid \DO(\overline{C}))$ is its \textit{counterfactual} probability in the hypothetic world as it is would have been had $C$ not occurred. One may easily verify that Equation (\ref{prob2}) holds in the three examples of Figure 1 by assuming that the switches are probabilistic and exogenous. In any case, under such circumstances, the causal attribution problem can thus be narrowed down to computing an estimate of the probabilities $\overline{p}$ and $p$. The latter only requires two experiments: a factual experiment $\DO(C)$ and a counterfactual one $\DO(\overline{C})$. Equation (\ref{prob}) then yields $\textrm{PN}, \textrm{PS}$ and $\textrm{PNS}$ from which a causal statement can be formulated.

Each three probability $\PS$, $\PN$ and $\PNS$ have different implications depending on the context. For instance, two perspectives can be considered: (i) the \textit{ex post} perspective of the plaintiff or the judge who asks ``does $C$ bear the responsibility of the event $E$ that did occur?''; and (ii) the \textit{ex ante} perspective of the planner or the policymaker who instead asks ``what should be done w.r.t. $C$ to prevent future occurrence of $E$?''.  It is $\textrm{PN}$ that is typically more relevant to context (i) involving legal responsibility, whereas $\textrm{PS}$ has more relevance for context (ii) involving policy elaboration. Both these perspectives could be relevant in the context of climate change, and it thus makes sense to trade them off. Note that $\PS$ and $\PN$ can be articulated with the conventional definition recalled in introduction. Indeed, the ``\textit{demonstration that the change is consistent with} (...)'' implicitly corresponds to the idea of sufficient causation, whereas ``(...) \textit{is not consistent with} (...)'' corresponds to that of necessary causation. The conventional definition therefore implicitly requires a high $\PS$ and a high $\PN$ to attribute a change to a given cause.

$\PNS$ may be precisely viewed as a probability which combines necessity and sufficiency. It does so in a conservative way since we have by construction that $\PNS\leq\textrm{min}(\PN,\PS)$. In particular, this means that a low $\PNS$ does not imply the absence of a causal relationship because either a high $\PN$ or a high $\PS$ may still prevail even when $\PNS$ is low. On the other hand, it presents the advantage that any statement derived from $\PNS$ asserting the existence of a causal link, holds both in terms of necessity and sufficiency. This property is thus prone to simplify causal communication, in particular towards the general public, since the distinction no longer needs to be explained. Therefore, establishing a high $\PNS$ may be considered as a suitable goal to evidence the existence of a causal relationship in a simple and straightforward way. In particular, the limiting case $\PNS=1$ corresponds to the fully deterministic, systematic and single-caused situation in Figure 1c --- i.e. undeniably the most stringent way in which one may understand causality.

%%%%%%%%%%%%%%%%%%%%%%%%%%%%%%%%%%
%%%%%%%%%%%%%%%%%%%%%%%%%%%%%%%%%%

\section{Probabilities of causation of climate change}

We now return to the question of interest: for a given forcing $f$ and an observed evolution of the climate system $y$, can $y$ be attributed to $f$? More precisely, what is the probability $\Proba(f\rightarrow y)$ that $f$ has caused $y$? We propose to tackle this problem by applying the causal counterfactual theory to the context of climate change, and more specifically, by using the three probabilities of causation $\PN$, $\PS$ and $\PNS$ recalled above. This Section shows that it can be done to a large extent similarly to the approach of \cite{HPONG15} for EA. 
In particular, as in EA, the crucial question to be answered as a starting point consists of narrowing down the definitions of the cause event $C$ and of the effect event $E$ associated to the question at stake --- where the word ``event'' is used here in its general mathematical sense of ``subset''.%In this respect, some differences between climate change attribution and extreme event attribution should be highlighted regarding the balance of necessary and sufficient causation evidence. Furthermore, a few useful specificities can also be introduced.

\subsection{Counterfactual setting}

For the cause event $C$, a straightforward answer is possible: we can follow the exact same approach as in EA by defining $C$ as ``presence of forcing $f$'' (i.e. the factual world that occurred) and $\overline{C}$ as ``absence of forcing $f$'' (i.e. the counterfactual world that would have occurred in the absence of $f$). Indeed, forcing $f$ can be switched on and off in numerical simulations of the climate evolution over the industrial period, as in the examples of Fig. 1  and as in standard EA studies. % thereby defining the factual world and the counterfactual one. 
Incidentally, the sample space $\Omega$ consists in the set of all possible climate trajectories in the presence and absence of $f$, including the observed one $y$. In other words, all forcings other than $f$ are held constant at their observed values as they are not concerned by the causal question. % which focuses exclusively on $f$. %While the above 

In practice and by definition, the factual runs of course always correspond to the historical experiment (HIST hereafter), using the Climate Model Intercomparison Project's (CMIP hereafter) terminology as described by \cite{Tay12}. The counterfactual runs are obtained from the same setting as historical but switching off the forcing of interest. For instance, if the forcing consists of the anthropogenic forcing then the counterfactual runs corresponds to the historicalNat (NAT hereafter) experiment i.e. $\Omega=\{ \textrm{HIST runs};\textrm{NAT runs}\}$. Likewise, if the forcing consists of the CO$_2$ forcing, then the counterfactual runs corresponds to the ``all except CO$_2$'' experiment. However, no such runs are available in CMIP5\\
(\verb|https://cmip.llnl.gov/cmip5/docs/historical_Misc_forcing.pdf| and Section 6 for discussion). Lastly, it is worth underlining that the historicalAnt experiment, which combines all anthropogenic forcings, thus corresponds to the counterfactual setting associated to the natural forcings. Therefore, runs from the historicalAnt experiment are relevant for the attribution of the natural forcings only, they are not relevant for the attribution of the anthropogenic forcings under the present counterfactual causal theory.

These definitions of $C$ and $\Omega$ %derive straightforwardly from the counterfactual approach buttressing the above causal theory, they have important implications 
have an important implication w.r.t. the design of numerical experiments in climate change attribution. In contrast with the design usually prevailing in D\&A (forcing $f$ only), the latter experiments are required to be counterfactual (i.e. all forcings except $f$). We elaborate further on this remark in Section 6.

\subsection{Balancing necessity and sufficiency}

To define the effect event $E$, %is sometimes not as straightforward as defining the cause event $C$. For this purpose, 
we propose to follow the same approach as in EA, where $E$ is usually defined based on an \textit{ad hoc} climatic index $Z$ exceeding a threshold $u$:
\begin{equation}
\label{defev}
E=\{Z\geq u\}
\end{equation}
%As discussed by \cite{HPONG15} and further highlighted in this Section, the choices of $Z$ and $u$ are to some extent arbitrary and they are far from neutral as they may hugely affect the balance between the probabilities of necessary and sufficient causation. 
Thus, defining $E$ implies choosing an appropriate variable $Z$ and threshold $u$ that reflect the focus of the question while keeping in mind the implications of the balance between the probabilities of necessary and sufficient causation. We now illustrate this issue and lay out some proposals to address it.

Consider the question ``\textit{Have anthropogenic CO$_2$ emissions caused global warming?}''. Following the above, the event ``\textit{global warming}'' may be loosely defined as a positive trend on global Earth surface temperature, i.e. $E=\{Z\geq0\}$ where $Z$ is the global surface temperature linear trend coefficient and the threshold $u$ is zero. In that case, $E$ nearly always occurs in the factual world ($p\simeq 1$) but it is also frequent in the counterfactual one ($\overline{p}$ medium) 
thus the emphasis is mostly on $\textrm{PS}$, i.e. on sufficient causation, while $\textrm{PN}$ and $\textrm{PNS}$ will have moderate values (Fig. 2b,e). But if global warming is more restrictively defined as a warming trend comparable to or greater than the observed trend, i.e. $E=\{Z\geq z\}$ where $u=z$ is the observed trend, then the event becomes nearly impossible in the counterfactual world ($\overline{p}\simeq 0$) but remains frequent in the factual one ($\overline{p}$ medium) thus the emphasis is on $\textrm{PN}$, i.e. on necessary causation, while the values of $\textrm{PS}$ and $\textrm{PNS}$ will this time be low. Therefore, the above two extreme definitions both yield a low $\PNS$. But under a more balanced definition of \textit{global warming} as a trend exceeding an intermediate value $u^*\in[0,z]$, then the event nearly always occurs in the factual world ($p\simeq 1$) and yet remains very rare in the counterfactual one ($\overline{p}\simeq 0$). Hence $\textrm{PNS}$ is then high: both necessary and sufficient causation prevail. %$\textrm{PNS}$ reaches a optimal value where . Thus, 
We propose to take advantage of this optimal value to define the event ``\textit{global warming}'' as the global trend index $Z$ exceeding the optimal threshold $u^*$ such that the amount of causal evidence, in a $\PNS$ sense, is maximized: 
\begin{equation}
\label{event1}
u^* = \textrm{argmax}_{u< z}\,\PNS(C\rightarrow \{Z\geq u\})
\end{equation}
where the condition $u< z$ insures that the event has actually occurred. %In words, $u^*$ defined by Equation (\ref{event1}) represents the level at which the best separation is obtained between the two distributions of $Z$ in the factual and counterfactual worlds. 
When used on real data (see Section 6), this approach yields a high value of $\PNS=0.95$ for the above question (Figure 2b,e).

%For the sake of highlighting a key distinction between attribution of climate change and attribution of weather event, l
Let us now consider the question ``\textit{Have anthropogenic CO$_2$ emissions caused the Argentinian heatwave of December 2013?}'' \citep{HVC15}. Here, the event can be defined as $E=\{Z\geq u\}$ where $Z$ is surface temperature anomaly averaged over an ad-hoc space-time window. Like in the previous case, the causal evidence agains shifts from necessary and not sufficient (Fig. 2a,d) when $u$ is equal to the observed value of the index $z=24.5^\circ$C (unusual event in both worlds but much more so in the counterfactual one) to sufficient and not necessary when $u$ is small (usual event in both worlds but much more so in the factual one). %In other words, adopting a less restrictive definition with a smaller threshold decreases the level of necessity but increases that of sufficiency. 
Like in the previous case, a possible approach here would be to balance both quantities by maximizing $\PNS$ in $u$ as in Equation (\ref{event1}). However, this would lead here to a substantially lower threshold which no longer reflects the rare and extreme nature of the event ``heatwave'' under scrutiny. Furthermore, this would yield a well-balanced, but pretty low level of causal evidence ($\PNS=0.35$). Thus maximizing $\PNS$ is not relevant here. Instead, maximizing $\PN$, even if that is at the expense of $\PS$, is arguably more relevant here since we are dealing with extreme events that are rare in both worlds, thereby inherently limiting the evidence of sufficient causation. This maximization corresponds to $u^* = \textrm{argmax}_{u< z}\,\PN(C\rightarrow \{Z\geq u\})$ which often yields the highest observed threshold $u=z$. %An immediate implication of this choice is also that by construction, only $\PN$ matters for EA, while $\PS$ and $\PNS$ are systematically low and can thus be considered as irrelevant to the analysis. 
Therefore, $\PN$ (i.e. the FAR) is an appropriate metric for the attribution of extreme events, and a high threshold $u$ matching with the observed value $z$ should be used in order to maximize it. %But only evidence of necessary causation can be obtained in that context, as sufficient causation is inherently small for extreme events. 
In contrast with extreme events, long term changes are prone to be associated with much powerful causal evidence that simultaneously involves necessary and sufficient causation, and may yield high values for $\PN$, $\PS$ and $\PNS$. $\PNS$ is thus an appropriate summary metric to consider for the attribution of climate changes, in agreement with D\&A guidance principles \citep{Guidance}. An optimal intermediate threshold can be chosen by maximizing $\PNS$.

\subsection{Building an optimal index}

In the above example where ``\textit{global warming}'' is the focus of the question, the variable of interest $Z$ to define the event can be considered as implicitly stated in the question, insofar as the term ``\textit{global warming}'' implicitly refers to an increasing trend on global temperature. However, in the context of climate change attribution, we often investigate the cause of ``an observed change $y$'' with no precise a priori regarding the characteristics of the change that are relevant w.r.t. causal evidencing. Furthermore, $y$ may be a large dimensional space-time vector. Thus the definition of the index $Z$ in this case is more ambiguous. 

We argue that in such a case, the physical characteristics of $y$ which are implicitly considered relevant to the causal question are precisely those which best enhance the existence of a causal relationship in a $\PNS$ sense. This indeed  corresponds to the idea of ``fingerprinting'' used thus far in climate change attribution studies (as well as in criminal investigations, hence the name): we seek a fingerprint, i.e. a distinctive characteristic of $y$ which would never appear in the absence of forcing $f$ (i.e. $\overline{p}\simeq 0$) but systematically does in its presence (i.e. $p\simeq 1$). If this characteristic shows up in observations, then the causal evidence is conclusive. %Likewise, in a police investigation, the causal question ``what is the probability that suspect $s$ has caused the observed crime scene $y$?'' is tackled by chasing the fingerprint of $s$ in the available data $y$. If found, then $s$ will likely be declared guilty beyond reasonable doubt. 
A fingerprint may thus be thought of as a characteristic which maximizes the gap between $p$ and $\overline{p}$ and thereby maximizes $\PNS$, by definition. % --- recalling that forcings are exogeneous and hence that $\PNS=p-\overline{p}$ (Eq. \ref{prob2}). 

As an illustration, \cite{MB13} focus on the attribution of changes in precipitation, and subsequently address the question ``\textit{Have anthropogenic forcing caused the observed evolution of precipitation at a global level?}''. Arguably, this study illustrates our point in the sense that it addresses the question by defining a \textit{fingerprint} index $Z$ which aims precisely at reflecting the features of the change in precipitation that are thought to materialize frequently (if not systematically) in the factual world and yet are expected to be rare (if not impossible) in the counterfactual one, based on physical considerations. %encompassing both dynamic and thermodynamic aspects. 
In practice, the index $Z$ defined by the authors consists of a non-dimensional scalar summarizing the main spatial and physical features of precipitation evolution w.r.t. dynamics and thermodynamics. The factual and counterfactual PDFs of $Z$ are then derived from the HIST and NAT runs respectively (Fig. 2c). %They %$(\overline{z}_0,\sigma_0)=(0.25,1.25)$ and $(\overline{z}_1,\sigma_1)=(1.75,1.25)$, are reproduced in Figure 4a. 
From these PDFs, one can easily obtain an optimal threshold $u^*$ for the precipitation index $Z$ by applying Equation (\ref{event1}) (Fig. 2f). This yields $\PNS=0.43$, i.e. anthropogenic forcings \textit{have about as likely as not caused the observed evolution of precipitation.}

A qualitative approach driven by physical considerations, such as the one of \cite{MB13}, is perfectly possible to define a fingerprint index $Z$ that aims at maximizing $\PNS$. However, a quantitative approach can also help in order to define $Z$ optimally, and to identify the features of $y$ that best discriminate between the factual and counterfactual worlds. Indeed, the qualitative, physical elicitation of $Z$ may be difficult when the joint evolution of the variables at stake is complex or not well-understood a priori. Furthermore, one may also wish to combine lines of evidence by treating several different variables at the same time in $y$ (i.e. precipitation and temperature, \cite{Yan16}). %In such situations, a mathematical approach may be relevant to help define $Z$. For the purpose of describing such an approach, 
Let us introduce the notation $Z=\phi(Y)$ where $Y$ is the space-time vectorial random variable of size $n$ which observed realization is $y$, and $\phi$ is a mapping from $\mathbb{R}^n$ to $\mathbb{R}$. Extending Equation (\ref{event1}) to the simultaneous determination of the optimal mapping $\phi^*$ and optimal threshold $u^*$, we propose the following maximization:
\begin{equation}
\label{event2}
(u^*,\phi^*) = \textrm{argmax}_{u<\phi(y), \,\phi \in \Phi}\,\,\PNS(C\rightarrow \{\phi(Y)\geq u\})
\end{equation}
In words, we thus propose to choose the value of the threshold, but also to choose the index $\phi$ among the set all possible indexes $\Phi$, so as to maximize $\PNS$.
%where $\Omega_{f,y}=\{E\subset\Omega_f\mid y\in E\}$; the event $E_{f,y}$ defined in Equation (\ref{event}) may thus be referred to as the \textit{optimal fingerprint} of $f$ w.r.t. $y$.
The event $E^* = \{\phi^*(Y)\geq u^*\}$ defined above in Equation (\ref{event2}) may thus be referred to as the \textit{optimal fingerprint} w.r.t. forcing $f$. The maximization performed in Equation (\ref{event2}) also suggests that our approach shares some similarity with the method of \cite{Yan16}, insofar as the variables of interest are in both cases selected mathematically by maximizing a criterion which is relevant for attribution (i.e. potential detectability in \cite{Yan16}, PNS in the present article), rather than by following qualitative, physics- or impact-oriented, considerations. %Assume for the time being that the problem of choosing a fingerprinting event $E_Y$ in the above $\PNS$-maximizing sense is solved. 

\section{Implementation under the standard framework}

We now turn to the practical aspects of implementing the approach described in Section 3 above, based on the observations $y$ and on climate model experiments. We detail these practical aspects in the context of the standard framework briefly recalled in Section 1, i.e. multivariate linear regression under a Gaussian setting. Note that the assumptions underlying the latter conventional framework could be challenged (e.g. pattern scaling description of model error and gaussianity). However, the purpose of this section is not to challenge these assumptions. It is merely to illustrate in detail how these assumptions can be used within the general causal framework proposed. Furthermore, the details of the mathematical derivation shown in this subsection can not be covered exhaustively here in order to meet the length constraint. However, some important steps of the derivation are described in Appendix A, and the complete details and justification thereof can be found in the references given in the text.

\subsection{Generalities}
The maximization of Equation (\ref{event2}) requires the possibility to evaluate the probabilities of occurrence $p$ and $\overline{p}$, in the factual and counterfactual world, of the event $\{\phi(Y)\geq u\}$, for any $\phi$ and $u$. %, based on the set of experiments $\Omega$. 
For this purpose, it is convenient to derive beforehand the factual and counterfactual PDFs of the random variable $Y$, denoted $\left[\, Y\mid f\,\right]$ and $\left[\, Y\mid \overline{f}\,\right]$ respectively. Assuming their two first moments are finite, we introduce:
\begin{equation} 
\label{gauss0}
\begin{array}{lll}
\Esp\left(Y\mid f\right) =\mu\,,&\,\,&\Var\left(Y\mid f\right) =\Sig\\
\Esp\left(Y\mid \overline f\right) =\overline{\mu}\,,&\,\,&\Var\left(Y\mid f\right) =\overline{\Sig}\\
\end{array}
\end{equation}
The means $\mu$ and $\overline{\mu}$ represent the expected response in the factual and counterfactual worlds; it is intuitive that their difference $\mu-\overline{\mu}$ will be key to the analysis. The covariances $\Sig$ and $\overline{\Sig}$ represent all the uncertainties at stake, they must be carefully determined based on additional assumptions. To avoid repetition in presenting these assumptions, we will detail them for the factual world only, but they will be applied identically in both worlds.

As recalled above, \textit{in situ} experimentation on the climate system is not accessible, thus we are left with \textit{in silico} experimentation as the only option. While the increasing realism of climate system models renders such an \textit{in silico} approach plausible, it is clear that modeling errors associated to their numerical and physical imperfections should be taken into account into $\Sig$. In addition, sampling uncertainty and observational uncertainty, which are commonplace sources of uncertainty in dealing with experimental results in an \textit{in situ} context as well, should also be taken into account. Finally, internal climate variability also needs to be factored. The latter four sources of uncertainty can be represented by decomposing $\Sig$ into a sum of four terms:
\begin{equation} 
\label{sigsum}
\begin{array}{lll}
\Sig = \mathbf C + \mathbf Q + \mathbf R + \mathbf S 
\end{array}
\end{equation}
where the component $\mathbf C$ represents climate internal variability; $\mathbf Q$ represents model uncertainty; $\mathbf R$ represents observational uncertainty; and $\mathbf S$ represents sampling uncertainty. Assumptions regarding the latter four sources of uncertainty are also key in the conventional Gaussian linear regression framework recalled in Section 1. We therefore propose to take advantage of some assumptions, data and procedures that have been previously introduced under the conventional framework, and adapt them to specify $\mu$, $\mathbf C$, $\mathbf Q$, $\mathbf R$ and $\mathbf S$. %and to subsederive the probabilities of causation of a climate change. 
The statistical model specification below can thus be viewed as an extension of developments under the conventional framework, in particular those exposed in \cite{Han16}. The various parameters and data involved, as well as their conditioning, are summarized in the hierarchical model represented in Figure 3, which we now describe.

\subsection{Model description}
%It is beyond the scope of the present paper to review these assumptions in detail. However, for the sake of illustrating our proposal, we describe one possibility which relies on the Gaussian assumption often used in the conventional framework. 

As in the conventional linear regression formulation recalled in Equation (\ref{regress}), %assumes the relationship $y = \sum_{i=1}^p \beta_i\,x_i+ \varepsilon$ where $\varepsilon$ is a Gaussian noise accounting for internal variability and observational error, and $\beta$ is a vector of scaling coefficients. Therefore, 
we assume that the random variable $Y$ is Gaussian with mean $\x\beta$ and covariance $\mathbf{C}+\mathbf{R}$:%accounting for model error. 
\begin{equation} 
\label{gauss1}
%\begin{array}{ll}
\left[\,Y\mid \beta,\x,\mathbf{C}, \mathbf{R}\,\right] = \mathcal{N}(\x\beta,\mathbf{C}+\mathbf{R})
%\end{array}
\end{equation}
In the conventional framework, climate models are assumed to correctly represent the response patterns $\x$ but to err on their amplitude. Recognizing that the scaling factors $\beta$ thereby aim at representing the error associated to models, we prefer to treat $\beta$ as a random variable instead of a fixed parameter to be estimated. The latter factors are also assumed to be Gaussian:
\begin{equation} 
\label{gauss2}
%\begin{array}{ll}
\left[\,\beta\mid\omega\,\right] =\mathcal{N}(e,\omega^2\mathbf I)
%\end{array}
\end{equation}
where we assume that the expected value of $\beta$ is $e=(1,...,1)'$, and $\omega$ is a scalar parameter which will be determined further in this Section. Combining Equations (\ref{gauss1}) and (\ref{gauss2}), it comes (Appendix A):
\begin{equation} 
\label{gauss3}
%\begin{array}{ll}
\left[\,Y\mid\mu, \x,\mathbf{C}, \mathbf{R},\omega\,\right] =\mathcal{N}(\mu,\mathbf{C}+\mathbf{R}+\omega^2\x\x')
%\end{array}
\end{equation}
where $\mu=\x e=\sum_{i=1}^p x_i$. Equation (\ref{gauss3}) thus shows that it is possible to translate the pattern scaling term $\x\beta$ from the mean of $Y$ to the covariance of $Y$. We believe such a mean-covariance translation is relevant here, since the pattern scaling assumption is meant to represent a source of uncertainty. Furthermore, the covariance $\mathbf Q$ associated to the latter source of uncertainty can be represented by the component $\omega^2\x\x'$, which results from the random scaling of the individual responses $x_1, x_2,...,x_p$. Furthermore, the expected value of $Y$, denoted $\mu$, is equal to the sum of the latter individual responses. Under the additivity assumption prevailing in the conventional framework, $\mu$ thus corresponds to the model response under the scenario where the $p$ forcings are present. Hence, $\mu$ can be obtained by estimating directly the combined response as opposed to estimating the individual responses $x_1, x_2,...,x_p$ one by one and summing them up. Such a direct estimation of $\mu$ is indeed advantageous from a sampling error standpoint, as will be made clear immediately below.

The PDF of $Y$ in Equation (\ref{gauss3}) involves three quantities $\mu$, $\x$ and $\mathbf C$ that needs to be estimated from a finite ensemble of model runs denoted $\E$, which of course introduces sampling uncertainty. %The latter sampling uncertainty can also conveniently be factored into the PDF of $Y$. 
Assuming independence among runs, it is straightforward to show that (Appendix A):
\begin{equation} 
\label{gauss4}
\begin{array}{ll}
\left[\,\mu\mid \mathbf{C}, \E\,\right] =\mathcal{N}(\widehat{\mu},\frac{1}{r}\mathbf{C})\,,&\,\,\,\,\left[\,x_i\mid \mathbf{C}, \E\,\right] \sim\mathcal{N}(\widehat{x}_i,\frac{1}{r_i}\mathbf{C})\,\,\,\textrm{for $i=1,...,p$}
\end{array}
\end{equation}
where $\widehat{x}_i$ is the ensemble average for the individual response $i$; $\widehat{\mu}$ is the ensemble average for the combined response; $r_i$ is the number of runs available for the individual response to forcing $i$; $r$ is the number of combined forcings runs. %$r_0$ the number of control runs; 
Combining Equations (\ref{gauss3}) and (\ref{gauss4}), and after some algebra, it comes (Appendix A):
\begin{equation} 
\label{gauss5}
%\begin{array}{ll}
\left[\,Y\mid \mathbf{C}, \mathbf{R},\E,\omega\,\right] =\mathcal{N}(\widehat{\mu},\mathbf{C}+\mathbf{R}+\omega^2\widehat{\x}\widehat{\x}'+\lambda\mathbf{C})
%\end{array}
\end{equation}
with $\lambda=1/r + \omega^2 \sum_i 1/r_i$, and where the sampling uncertainty $\mathbf S$ on the responses $\mu$ and $\x$ thus corresponds to the term $\lambda\mathbf{C}$.
 On the other hand, the internal variability component $\mathbf C$ also has to be estimated from the $r_0$ preindustrial control runs, which introduces additional sampling uncertainty. The sampling uncertainty on $\mathbf C$ can be treated by following the approach of \cite{Han16}, which introduces an Inverse Wishart conjugate prior for $\mathbf C$. This leads an Inverse Wishart posterior for $\mathbf C$ which has the following expression (Appendix A):
\begin{equation} 
\label{wishart}
%\begin{array}{ll}
\left[\,\mathbf{C}\mid \E\,\right] = \mathcal{IW}(\widehat{\mathbf{C}},\widehat{\nu})
%\end{array}
\end{equation}
where the estimated covariance $\widehat{\mathbf{C}}$ consists of a so-called shrinkage estimator:
\begin{equation} 
\label{wishart2}
\begin{array}{ll}
\widehat{\mathbf{C}} = \widehat{a} \widehat{\boldsymbol{\Delta}}+ (1-\widehat{a})\widehat{\Om}\\
\end{array}
\end{equation}
where $\widehat{\Om}$ is the empirical covariance of the control ensemble; $\boldsymbol{\Delta}$ is a shrinkage target matrix taken here to be equal to $\textrm{diag}(\widehat{\Om})$ i.e. $\widehat{\boldsymbol{\Delta}}_{ii}=\widehat{\Om}_{ii}$ and $\widehat{\boldsymbol{\Delta}}_{ij}=0$ for $i\neq j$; the shrinkage intensity $\widehat{a}$ is obtained from the marginal likelihood maximization described in \cite{HN14}; and $\widehat{\nu}= 2+ r_0/(1-\widehat a)$.

Combining Equations (\ref{gauss5}) and (\ref{wishart}), and after some algebra and an approximation, it comes (Appendix A):
\begin{equation} 
\label{student}
%\begin{array}{ll}
\left[\,Y\mid \E,\omega,\sigma\,\right] = \mathcal{S}t(\widehat{\mu},\sigma^2 \mathbf I+\omega^2\widehat{\x}\widehat{\x}'+(1+\lambda)\widehat{\mathbf{C}},\widehat{\nu})
%\end{array}
\end{equation}
where we adopted the simplified parametric form $\mathbf R=\sigma^2\mathbf I$ for the covariance of observational error, and where $\mathcal{S}t(\mu,\Sig,\nu)$ is the multivariate $t$ distribution with mean $\mu$, variance $\Sig$ and $\nu$ degrees of freedom. Equation (\ref{student}) implies that taking into account the sampling uncertainty on $\mathbf C$ does not affect the total variance of $Y$. Instead, it only affects the shape of the PDF of $Y$, which has thicker tails than the Gaussian distribution. With these parameterizations, our model for $Y$ is thus a parametric Student $t$ model with two parameters $(\sigma, \omega)$. 

After computing the estimators $\widehat{\mu}$, $\widehat{\x}$, $\widehat{\mathbf{C}}$ and $\widehat{\nu}$ using the ensemble of model experiments as described above, the parameters  $(\sigma, \omega)$ are estimated by fitting the above model to the observation $y$ based on likelihood maximization. The log-likelihood of the model has the following expression:
\begin{equation} 
\label{llk}
\begin{array}{ll}
\ell(\sigma, \omega; y)= &-\frac{1}{2}\log\vert(1+\lambda)\widehat{\mathbf{C}}+\sigma^2 \mathbf I+\omega^2\widehat{\x}\widehat{\x}'\vert\\
&-\frac{1}{2}(\widehat\nu+n)\log\left(1+\frac{1}{\widehat\nu-2}(y-\widehat{\mu})'\left((1+\lambda)\widehat{\mathbf{C}}+\sigma^2 \mathbf I+\omega^2\widehat{\x}\widehat{\x}'\right)^{-1}(y-\widehat{\mu})\right)
\end{array}
\end{equation}
The estimators $\widehat{\sigma}$ and $\widehat{\omega}$ are then obtained numerically using a standard maximization algorithm (e.g. gradient descent). With $\widehat{\mu}$ being obtained from factual runs (i.e. HIST runs) and $\widehat{\x}$ containing all the forcings including $f$, this procedure thus yields the PDF of $Y$ in the factual world: 
\begin{equation} 
\label{factPDF}
\begin{array}{ll}
\left[\, Y\mid f\,\right] = \mathcal{S}t(\widehat{\mu},\widehat{\Sig},\widehat{\nu})\\
\widehat{\Sig}=(1+\widehat{\lambda})\widehat{\mathbf{C}}+\widehat{\sigma}^2 \mathbf I+\widehat\omega^2\widehat{\x}\widehat{\x}'
\end{array}
\end{equation}
Next, to obtain $\left[\, Y\mid \overline{f}\,\right]$, one simply needs to change the mean $\widehat{\mu}$ to $\widehat{\overline{\mu}}$ obtained as the ensemble average for the counterfactual experiment ``all forcings except $f$''. Some changes also need to be made regarding the covariance. Indeed, since forcing $f$ is absent in the counterfactual world, the model error covariance component $\widehat\omega^2\widehat x_f\widehat x_f'$, corresponding to the random scaling of the response $\widehat x_f$ to forcing $f$, must be taken out of the covariance. Furthermore, if the number of counterfactual runs $\overline r$ differ from the number of factual runs $r$, the sampling uncertainty $\widehat{\mathbf C}/r$ associated to estimating $\mu$ also has to be modified. The PDF of $Y$ in the counterfactual world can thus be written:
\begin{equation} 
\label{cfactPDF}
\begin{array}{ll}
\left[\, Y\mid \overline{f}\,\right] = \mathcal{S}t(\widehat{\overline\mu},\widehat{\overline\Sig},\widehat{\nu})\\
\widehat{\overline\Sig}=\widehat{\Sig}-\widehat\omega^2\widehat x_f\widehat x_f'+(\frac{1}{\overline r}-\frac{1}{r})\widehat{\mathbf C}
\end{array}
\end{equation}
As noted above, when $f$ is anthropogenic forcing, the counterfactual experiment NAT is usually available in CMIP runs, allowing for a straightforward derivation of $\widehat{\overline\mu}$. But for other forcings, by the design of CMIP experiments, counterfactual runs are usually not available. A possible workaround then consists in applying the additivity assumption to approximate $\widehat{\overline{\mu}}$ with $\widehat{\mu}-\widehat{x}_f$. For instance, if CO$_2$ is the forcing of interest, the counterfactual response to all forcings except CO$_2$ emissions can be approximated by subtracting the CO$_2$ individual response $x_f$ from the all forcings response. However in that case, the sampling uncertainty term $\widehat{\mathbf C}/r_f$ corresponding to the estimation of $\widehat{x}_f$ must be added to the covariance $\widehat{\overline\Sig}$.

%but may be regularized or modified to account for model's limitations \citep{RAP09,HN14}. The observational error component $\mathbf R$ may be obtained from data products providing an ensemble of values reflecting uncertainty \citep{Mo12}, or may require specific assumptions as well. In any case, as the ongoing discussion on the warming hiatus suggests \citep{Mee13,Karl15}, both climate variability and observational error appear to significantly modify the signal, even when the latter is averaged globally. Accordingly, assumptions on $\mathbf C$ and $\mathbf R$ are important as they are prone to significantly modify $\Proba(f\rightarrow y)$. 

\subsection{Derivation of the probabilities of causation}

With the two PDFs of $Y$ in hand, an approximated solution to the maximization of Equation (\ref{event2}) can be conveniently obtained by linearizing $\phi$, yielding a closed mathematical expression for the optimal index $\phi^*(Y)$:
\begin{equation} 
\label{optphi}
\begin{array}{ll}
\phi^*(Y) = (\widehat{\mu}-\widehat{\overline{\mu}})'\widehat{\Sig}^{-1}Y
\end{array}
\end{equation}
Equation (\ref{optphi}) is a well known result of Linear Discriminant Analysis (LDA) \citep{McLachlan04}. Details of approximations made and of the mathematical derivation of Equation (\ref{optphi}) are given in Appendix B. The optimal index $Z^*=\phi^*(Y)$ can thus be interpreted as the projection of $Y$ onto the vector $\widehat{\Sig}^{-1}(\widehat{\mu}-\widehat{\overline{\mu}})$ which will be denoted $\phi^*$ hereinafter, i.e. $\phi^*(Y)\equiv \phi^{*'}Y$.

To obtain $\PNS$, we then need to derive the factual and counterfactual CDFs of $Z=\phi^*(Y)$, denoted $G$ and $\overline G$ respectively. Since no closed form expression of these CDFs is available, we simulate realizations thereof. Drawing two samples of $N$ random realizations of $Y$ from the Student $t$ distributions $\left[\, Y\mid f\,\right]$ and $\left[\, Y\mid \overline{f}\,\right]$ is straightforward, by treating the Student $t$ as a compound Gaussian--Chi-squared distribution. Samples of $Z$ are then immediately obtained by projecting onto $\phi^*$ and the desired CDFs can be estimated using the standard kernel smoothing estimator \citep{BA97}, yielding $\widehat{G}(u)$ and $\widehat{\overline G}(u)$ for all $u\in\mathbb R$. Finally, $\PNS^*$ follows as:
\begin{equation}
\label{pns}
\begin{array}{ll}
\PNS^* = \widehat{\overline G}(u^*)-\widehat{G}(u^*)
\end{array}
\end{equation}
and:
\begin{equation}
\label{pn_ps}
%\begin{array}{ll}
\PN^* = 1-\frac{1-\widehat{\overline G}(u^*)}{1-\widehat{G}(u^*)}\,,\,\,\,\,\,\,\,\PS^* = 1-\frac{\widehat{G}(u^*)}{\widehat{\overline G}(u^*)}
%\end{array}
\end{equation}
where $u^*=\textrm{argmax}_{u< z}\, \{\widehat{\overline G}(u)-\widehat{G}(u)\}$.

\subsection{Reducing computational cost}

When the dimension of $y$ is large, the above described procedure can become prohibitively costly if applied straightforwardly, due to the necessity to derive the inverse and determinant of $\widehat\Sig$ at several steps of the procedure. However, the computational cost of these operations can be drastically reduced by applying the Sherman-Morrison-Woodbury formula \citep{Woo50}, which states that the inverse of a low rank correction of some matrix can be computed by doing a low rank correction to the inverse of the original matrix. Omitting the notation $\widehat{.}$ for more clarity, we have:
\begin{equation}
\label{sherman1}
\begin{array}{ll}
\Sig^{-1}%&=\left((1+\lambda)\mathbf{C}+\sigma^2 \mathbf I+\omega^2\x\x'\right)^{-1}\\
%&=\left(\mathbf A+\omega^2\x\x'\right)^{-1}\\
=\mathbf A^{-1}-\omega^2\mathbf A^{-1}\x(\mathbf I +\omega^2\x'\mathbf A^{-1}\x)^{-1}\x'\mathbf A^{-1}\\
\end{array}
\end{equation}
where $\mathbf A=(1+\lambda)\mathbf{C}+\sigma^2 \mathbf I$ can be inverted using the same formula:
\begin{equation}
\label{sherman2}
\begin{array}{ll}
\mathbf A^{-1}%&=\left((1+\lambda)\mathbf{C}+\sigma^2 \mathbf I\right)^{-1}\\
%&=\left((1+\lambda)a\boldsymbol{\Delta}+\sigma^2 \mathbf I+\frac{1}{r_0}(1+\lambda)(1-a)\x_0\x_0'\right)^{-1}\\
%&=\left(\mathbf B+\frac{1}{r_0}(1+\lambda)(1-a)\x_0\x_0'\right)^{-1}\\
=\mathbf B^{-1}-\frac{1}{r_0}(1+\lambda)(1-a)\mathbf B^{-1}\x_0(\mathbf I +\frac{1}{r_0}(1+\lambda)(1-a)\x_0'\mathbf B^{-1}\x_0)^{-1}\x_0'\mathbf B^{-1}\\
\end{array}
\end{equation}
where $\mathbf B=(1+\lambda)a\boldsymbol{\Delta}+\sigma^2 \mathbf I$. Likewise, we apply the Sylvester formula \citep{Syl51} twice to compute the determinant of $\Sig$:
\begin{equation}
\label{sylvester1}
\begin{array}{ll}
\vert\Sig\vert%&=\vert(1+\lambda)\mathbf{C}+\sigma^2 \mathbf I+\omega^2\x\x'\vert\\
%&=\vert\mathbf A+\omega^2\x\x'\vert\\
&=\vert\mathbf A\vert\,.\,\vert\mathbf I+\omega^2\x'\mathbf A^{-1}\x\vert\\
&=\vert \mathbf B\vert\,.\,\vert\mathbf I +\frac{1}{r_0}(1+\lambda)(1-a)\x_0'\mathbf B^{-1}\x_0\vert\,.\,\vert\mathbf I+\omega^2\x'\mathbf A^{-1}\x\vert\\
\end{array}
\end{equation}
Independently of $n$, the matrices $\mathbf I +\omega^2\x'\mathbf A^{-1}\x$ is of size $p$, $\mathbf I +\frac{1}{r_0}(1+\lambda)(1-a)\x_0'\mathbf B^{-1}\x_0$ is of size $r_0$, and $\mathbf B$ is diagonal. Obtaining their inverse and determinant is therefore computationally cheap. Hence, the inverse and determinant of $\Sig$ can be obtained at a low computational cost by applying first Equation (\ref{sherman2}) to determine $\mathbf A^{-1}$ and second Equations (\ref{sherman1}) and (\ref{sylvester1}).

\section{Illustration on temperature change}

Our methodological proposal is applied to the observed evolution of Earth's surface temperature during the $20^{\textrm{th}}$ century, with the focus being restrictively on the attribution to anthropogenic forcings. More precisely, $y$ consists of a spatial-temporal vector of size $n=54$ which contains the observed surface temperatures averaged over 54 time-space windows. These windows are defined at a coarse resolution: Earth's surface is divided into 6 regions of similar size (3 in each hemisphere) while the period 1910-2000 is divided into 9 decades. The decade 1900-1910 is used as a reference period, and all values are converted to anomalies w.r.t. the first decade. The HadCRUT4 observational dataset \citep{Mo12} was used to obtain $y$. With respect to climate simulations, the runs of the IPSL-CM5A-LR model \citep{Duf12} for the NAT, ANT, HIST and PIcontrol experiments were used (see Appendix C for details) and converted to the same format as $y$ after adequate space-time averaging.

Following the procedure described in Section 4, we successively derived the estimated factual response $\widehat\mu$ using the $r$ HIST runs; the estimated counterfactual response $\widehat{\overline{\mu}}$ using the $\overline r$ NAT runs; the estimated individual responses $x_1$ and $x_2$ using the $r_1$ NAT runs and $r_2$ ANT runs respectively ($p=2$ and $\x=(x_1,x_2)$); the estimated covariance $\widehat{\mathbf C}$ from the $r_0$ PIcontrol runs. Then, we derived $\widehat{\sigma}$ and $\widehat{\omega}$ by likelihood maximization, to obtain  $\widehat{\Sig}$ and $\widehat{\overline\Sig}$. 

An assessment of the relative importance of the four components of uncertainty was obtained by deriving the trace of each component (i.e. the sum of diagonal terms) normalized to the trace of the complete covariance. Climate variability is found to be the dominant contribution, followed by model uncertainty, observational uncertainty and sampling uncertainty (not shown). %Model uncertainty is almost negligible in the counterfactual world because the NAT response is flat. Note that t
The split between model and observational uncertainty is to some extent arbitrary as we have no objective way to separate them based only on $y$, i.e. the model could be equivalently formulated as $\mathbf Q=\omega^2\x\x'+(1-\alpha)\sigma^2\mathbf I$ and $\mathbf R=\alpha\sigma^2\mathbf I$. An objective separation would require an ensemble representing observational uncertainty, allowing for a preliminary estimation of $\mathbf R$.

The optimal vector $\phi^*$, designed to capture the space-time patterns that best discriminate the HIST evolution and the NAT one, was then obtained from Equation (\ref{optphi}). To identify which features of $Y$ are captured by this optimal mapping, the coefficients $(\phi^*_1,...,\phi_n^*)$ were averaged spatially and temporally, and were plotted in Figure 4.  Firstly, it can be noted that the coefficients' global average $\langle\phi^*\rangle= \sum_{i=1}^n  \phi^*_i$ is large and positive: a major discriminant feature is merely global mean temperature, as expected. Secondly, the coefficients also exhibit substantial variation around their average $\langle\phi^*\rangle$ in both space and time. Spatial variations of $\phi^*$ unsurprisingly suggest that, beyond global mean temperature, other discriminant features include the warming contrast prevailing between the two hemispheres and/or between low and high latitudes (the low resolution prevent from a clear separation), as well as between ocean and land (Fig. 4a). Temporal variations of $\phi^*$ suggest that discriminant features includes the linear trend increase as expected, but also higher order temporal variations (Fig. 4b).

The PDFs of the optimal index $Z=\phi^{*'}Y$ were derived, and are plotted in Figure 5, together with $\PNS$ as a function of the threshold $u$. The maximum of $\PNS$ determines the desired probability of causation:
\begin{equation}
\label{pns_res}
\begin{array}{ll}
\Proba(\textrm{ANT}\rightarrow y) = 0.9999
\end{array}
\end{equation}
In IPCC terminology, this would mean that anthropogenic forcings \textit{have virtually certainly caused the observed evolution of temperature}, according to our approach. More precisely, the probability that the observed evolution of temperature is not caused by anthropogenic forcings is one in then thousands (1:10,000) instead of one in twenty (1:20). Therefore, the level of causal evidence found here is substantially higher than the level assessed in the IPCC report. This discrepancy will be discussed in Section 6. 

Before digging into this discussion, it is interesting to assess the relative importance of the ``trivial'' causal evidence coming from the global increase in temperature, and of the less obvious causal evidence coming from space-time features captured by $\phi^*$. For this purpose, we merely split $\phi^*$ into the sum of a global average contribution $ \sum_{i=1}^n \langle\phi^*\rangle Y_i$ and of the remaining variations $\sum_{i=1}^n (\phi_i^*-\langle\phi^*\rangle) Y_i$. The PDFs of the resulting indexes are plotted in Figure 5a,b. Their bivariate PDF can also be visualized with the scatterplot of Figure 5c. The following two probabilities of causation are obtained:
\begin{equation}
\label{pns_res}
\begin{array}{ll}
\Proba(\textrm{ANT}\rightarrow \langle y\rangle) = 0.9781\\
\Proba(\textrm{ANT}\rightarrow y-\langle y\rangle) = 0.9994
\end{array}
\end{equation}
where $\langle y\rangle$ refer to the globally averaged evolution and $y-\langle y\rangle$ refer to other features of evolution. Therefore, while the globally averaged warming provides alone a substantial level of evidence (i.e. $\Proba(\textrm{ANT}\rightarrow \langle y\rangle) = 0.9781$), these results suggest that the overwhelmingly high overall evidence (i.e. $\Proba(\textrm{ANT}\rightarrow y) = 0.9999$) is primarily associated to non-obvious space-time features of the observed temperature change. 

\section{Discussion}

\subsection{Comparison with previous statements}

The probabilities of causation obtained by using our proposal may depart from the levels of uncertainty asserted by the latest IPCC report, and/or by previous work. For instance, when $y$ corresponds to the evolution of precipitation observed over the entire globe during the satellite era (1979-2012), we have shown in Section 3 that, using the dynamic-thermodynamic index built by \cite{MB13}, the associated probability of causation $\Proba(\textrm{ANT}\rightarrow y)$ is found to be 0.43. This probability is thus significantly lower than the one implied by the claim made in this study that ``\textit{the changes in precipitation observed in the satellite era are likely to be anthropogenic in nature}'' wherein ``\textit{likely}'' implicitly means $\Proba(\textrm{ANT}\rightarrow y)\geq0.66$.
%\textit{external influences are responsible for the observed precipitation changes.}

In contrast with the situation prevailing for precipitation, when $y$ corresponds to the observed evolution of Earth's surface temperature during the $20^{\textrm{th}}$ century, and in spite of using a very coarse spatial resolution, we found a probability of causation $\Proba(\textrm{ANT}\rightarrow y)=0.9999$ which considerably exceeds the $0.95$ probability implied by the latest IPCC report. Such a gap deserves to be discussed. 

Firstly, the probability of causation defined in our approach is of course sensitive to the assumptions that are made on the various sources of uncertainty, all of which are here built into $\Sig$. Naturally, increasing the level of uncertainty, for instance through an inflation factor applied to $\Sig$, reduces the probability of causation (Figure 6). In the present illustration, uncertainty needs to inflated by a factor 2.4 to obtain $\Proba(\textrm{ANT}\rightarrow y)=0.95$ in agreement with the IPCC statement. Therefore, a speculative explanation for the gap is that experts may be adopting a conservative approach by implicitly inflating uncertainty, although not explicitly, perhaps in an attempt to account for additional sources of uncertainty that are not well known. In the present case, such an inflation should amount to 2.4 to explain the gap. This number is arguably too high to provide a satisfactory standalone explanation, yet overall, such a conservativeness may partly contribute to the discrepancy when it comes to temperature. %However, no such conservativeness seems to be at play w.r.t. precipitation. 
In any case, this highlights the need for a more explicit and consistent use of conservativeness --- if any.

Besides the effect of inflating the individual variances, it is important to note that the probability of causation may also be greatly reduced when the correlation coefficients of the covariance $\Sig$, whether spatial or temporal, are inflated. This less straightforward effect can be explained by the fact that higher correlations imply greater spatial and temporal coherence of the noise, which is thus more prone to confounding with a highly coherent signal, and thereby reduces the probability of causation. Conservativeness may thus be associated to an inflation of correlations, in addition to an inflation of variances.

Another possible explanation for the discrepancy is more technical. Many previous attribution studies buttressing the IPCC statement regarding temperature, are based on an inference method for the linear regression model of Equation (\ref{regress}) which is not optimal w.r.t. maximizing causal evidence --- despite of it being often referred to as ``\textit{optimal fingerprinting}''. More precisely, the inference on the scaling factors $\beta$ and the associated uncertainty quantification, are obtained by projecting the observation $y$ as well as the patterns $\x$ onto the leading eigenvectors of the covariance $\mathbf C$ associated to climate internal variability. Such a projection choice sharply contrasts with the projection used in our approach, which is indeed performed onto the vector $\phi^*= \Sig^{-1}(\mu-\overline{\mu})$. Denoting $(\vv_1,...,\vv_n)$ the eigenvectors of $\Sig$ and $(\lambda_1,...,\lambda_n)$ the corresponding eigenvalues, the expression of $\phi^*$ can be written:
\begin{equation} 
\label{optphi3}
%\begin{array}{ll}
\phi^* =\sum_{k=1}^n \frac{\langle \vv_k\mid \mu-\overline{\mu}\rangle}{\lambda_k}\,.\,\vv_k\\
%&=\sum_{k=1}^r \frac{\langle \vv_k\mid \mu-\overline{\mu}\rangle}{\lambda_k}\,.\,\vv_k+\sum_{k=r+1}^n \frac{\langle \vv_k\mid \mu-\overline{\mu}\rangle}{\lambda_k}\,.\,\vv_k\\
%\phi^*&\simeq(\mathbf I - \mathbf V_r  \mathbf V_r')(\mu-\overline{\mu})
%\end{array}
\end{equation}
%
%\begin{equation} 
%\label{optphi3b}
%\begin{array}{ll}
%\sum_{k=r+1}^n \frac{\langle \vv_k\mid \mu-\overline{\mu}\rangle}{\lambda_k}\,.\,\vv_k&\simeq\sum_{k=r+1}^n \langle \vv_k\mid \mu-\overline{\mu}\rangle\,.\,\vv_k\\
%&=(\mu-\overline{\mu})-\sum_{k=1}^r \langle \vv_k\mid \mu-\overline{\mu}\rangle\,.\,\vv_k\\
%\end{array}
%\end{equation}
%
%$\mathbf V_r = (\vv_1,...,\vv_r)$
%
Equation (\ref{optphi3}) shows that projecting onto $\phi^*$ does not emphasize the leading eigenvectors of $\Sig$, in contrast to the aforementioned standard projection. Instead, it emphasizes the eigenvectors that simultaneously present a low eigenvalue $\lambda_k$ and a strong alignment with the contrast between the two worlds $\mu-\overline{\mu}$. As a matter of fact, the ratio $\langle \vv_k\mid \mu-\overline{\mu}\rangle/\lambda_k$ can be interpreted as the signal-to-noise ratio associated to the eigenvector $\vv_k$, where the eigenvalue $\lambda_k$ quantifies the magnitude of the noise and $\langle\vv_k\mid \mu-\overline{\mu}\rangle$ that of the causal signal. Projecting onto $\phi^*$ thus maximizes the signal-to-noise ratio. In contrast, since $\mathbf C$ is a large contribution to $\Sig$ (the dominant one in our illustration), a projection onto the leading eigenvectors of $\mathbf C$ naturally tends to amplify the noise, and to partly hide the relevant causal signal $\mu-\overline{\mu}$.

In order to assess whether or not these theoretical remarks hold in practice, we revisited our illustration and quantified the impact on $\Proba(\textrm{ANT}\rightarrow y)$ of using such a projection onto the leading eigenvectors of $\mathbf C$. For this purpose, after computing the projection matrix $\mathbf P$ on the ten leading eigenvectors of $\mathbf C$, our procedure was applied identically, but this time using the projected vector $\phi^+ = \mathbf P\phi^*$. Results are shown in Figure 7, again after splitting the contribution of global mean change and patterns of change. Unsurprisingly, the probability of causation associated to the global mean change remains unmodified at 0.978. However, the probability of causation associated to the space-time features of warming drops from 0.9994 to 0.92. %It almost does not contribute to the final value obtained, $\Proba(\textrm{ANT}\rightarrow y)=0.98$. 
Indeed, the features that most discriminate the two worlds, and are summarized in $\phi^*$, do not align well with the leading eigenvectors of $\mathbf C$. They are thus incompletely reflected by the projected vector $\phi^+$ (Figure 8). Furthermore, the uncertainty of the resulting index $Z^+=\phi^{+'}Y$ is high by construction, as the magnitude of climate variability is maximized when projecting onto its leading modes. This further contributes to reducing $\Proba(\textrm{ANT}\rightarrow y)$ to 0.992.

\subsection{Counterfactual experiments}
Our methodological proposal has an immediate implication w.r.t. the design of standardized CMIP experiments dedicated to D\&A: a natural option would be to change the present design ``forcing $f$ only'' into a counterfactual design ``all forcings except $f$''. Indeed, $\Proba(f\rightarrow y)$ is driven by the difference $\mu-\overline{\mu}_f$ between the factual response $\mu$ (i.e. historical experiment) and the counterfactual response $\overline{\mu}_f$ (i.e. all forcings except $f$ experiment).  Under the assumption that forcings do not interact with one another and that the combined response matches with the sum of the individual responses, the difference $\mu-\overline{\mu}_f$ coincides with the individual response $x_f$ (i.e. forcing $f$ only experiment). While this hypothesis is well established for temperature at large scale \citep{Gill04}, it appears to break down for other variables (e.g. precipitation, \citep{Sh13}) or over particular regions (e.g the Southern extratropics, \citep{Morg14}) where forcings appear to significantly interplay. Such a lack of additivity would inevitably damage the results of the causal analysis. It is thus important in our view to better understand the domain of validity of the forcing-additivity assumption and to evaluate the drawbacks of the present ``one forcing only'' design versus its advantages. Such an analysis does require ``forcing $f$ only'' experiments, but also ``all forcings except $f$'' experiments in order to allow for comparison. This effort would hence justify including in the DAMIP set of experiments an ``all forcings except $f$'' experiment --- which is presently absent even in the lowest priority tier thereof --- at least for the most important forcings such as anthropogenic CO$_2$.  %Counterfactual analysis is possible under CMIP5 but it is confined to the NAT and ANT runs. Indeed, the NAT experiment can be considered as above as as a simulation of the world with all forcings except anthropogenic ones and thereby be used to counterfactually evidence the latter, instead of factually evidencing natural forcings as is presently done under the OF approach. Generalizing this approach to all forcings would in our view be a step forward for D\&A.

\subsection{Benchmarking high probabilities}
Section 5 showed that the proposed approach may sometimes yield probabilities of causation that are very close to one. How can we communicate such low levels of uncertainty? This question arises insofar as the term ``virtual certainty''  applies as soon as $\PNS$ exceeds $0.99$ under the current IPCC language. Thus, this terminology would be unfit to express in words a $\PNS$ increase from $0.99$ to $0.9999$, say --- even though such an increase corresponds to a large reduction of uncertainty by a factor one hundred. One option to address this issue is to use instead the uncertainty terminology of theoretical physics, in which a probability is translated into an exceedance level under the Gaussian distribution, measured in numbers of $\sigma$ from the mean (where $\sigma$ denotes standard deviation), i.e. $F^{-1}(\PNS)\sigma$ with $F$ the CDF of the standard Gaussian distribution. Under such terminology, ``virtual certainty'' thus corresponds to a level of uncertainty of 2.3$\sigma$, while $\Proba(\textrm{ANT}\rightarrow y)=0.9999$ found in Section 5 reaches 3.7$\sigma$. It is interesting to note that the level of uncertainty officially requested in theoretical physics to corroborate a discovery as such --- e.g. the existence of the Higgs Boson --- is 5$\sigma$. By applying such standards, one may actually consider that $\Proba(\textrm{ANT}\rightarrow y)=0.9999$ is still too low a probability to corroborate that human influence has indeed been the cause of the observed warming. Whether or not such standards are relevant in the particular context of climate change --- which relates to defining the proper level of aversion to false discovery suitable in that context --- are discutable matters. In any case, increasing $\Proba(\textrm{ANT}\rightarrow y)$ beyond the ``virtual certainty'' threshold of $0.99$ by building more evidence into the analysis, is possible and may still be considered as a relevant research goal.

\subsection{Alternative assumptions}

The mathematical developments of Section 4 are but an illustration of how our proposed causal approach, as framed in Section 3, can be implemented when one uses the conventional assumptions of pattern scaling and gaussianity associated to the standard linear regression setting. In that sense, Section 4 thus shows that the proposed causal framing is perfectly compatible with the conventional linear regression setting: it should be viewed as an extension of, rather than an alternative to, the latter setting. Nevertheless, it is important to underline that the application of the causal framework of Section 3 is by no means restricted to the conventional linear regression setting. One may for instance challenge some aspects of the latter, e.g. the pattern scaling description of model error, and formulate an alternative parameterization of the covariance $\Sig$. This does not affect the relevance of the maximization of Equation (\ref{event2}), which can be implemented similarly.

\subsection{Attribution as a classification problem}

Lastly, it should be noted that the maximization of Equation (\ref{event2}) can be viewed as a binary classification problem. Indeed, as illustrated in Figure 5, solving Equation (\ref{event2}) is equivalent to building a function of observations which allows to optimally discriminate between two ``classes'': the factual class and the counterfactual class. Under this perspective, PNS is related to the $\%$ of correct classification decisions made by the classifier and is thus a measure of its skill.

Viewing the fingerprinting index $\phi^*$ as a classifier offers a fruitful methodological angle in our opinion. Indeed, classification is a classic and widespread problem in statistics and machine learning for which numerous solutions are readily available. For instance, under the restrictive assumptions of Section 4 --- among which the Gaussian assumption --- one obtains a linear classifier under a closed form expression, which is well known in Linear Discriminant Analysis \citep{McLachlan04}. But more recent developments have focused on the non Gaussian situations where a nonlinear classifier is more suitable, and can be formulated for instance using as diverse approaches as random forests, support vector machine or neural nets \citep{Alp10}. Testing such approaches for the present attribution problem certainly offers promise. However, the difficulty to physically interpret such complex classifiers represent a challenge in such approaches. 

\section{Summary and conclusion}
We have introduced an approach for deriving the probability that a forcing has caused a given observed change. The proposed approach is anchored into causal counterfactual theory \citep{Pearl09} which has been introduced recently in the context of EA. We argued that these concepts are also relevant, and can be straightforwardly extended to the context of climate change attribution. For this purpose, and in agreement with the principle of \textit{fingerprinting} applied in the conventional D\&A framework, a trajectory of change is converted into an event occurrence defined by maximizing the causal evidence associated to the forcing under scrutiny. Other key assumptions used in the conventional D\&A framework, in particular those related to numerical models error, can also be adapted conveniently to this approach. Our proposal thus allows to bridge the conventional framework with the standard causal theory, in an attempt to improve the quantification of causal probabilities. Our illustration suggested that our approach is prone to yield a higher estimate of the probability that anthropogenic forcings have caused the observed temperature change, thus supporting more assertive causal claims.

\begin{acknowledgment}
We gratefully acknowledge helpful comments by Aur\'elien Ribes and three anonymous reviewers. This work was supported by the French Agence Nationale de la Recherche grant DADA (AH, PN), and the grants LEFE-INSU-Multirisk, AMERISKA, A2C2, and Extremoscope (PN). The work of PN was completed  during his visit at the IMAGE-NCAR group in Boulder, CO, USA.
\end{acknowledgment}

% Use appendix}[A], {appendix}[B], etc. etc. in place of appendix if you have multiple appendixes.

\ifthenelse{\boolean{dc}}
{}
{\clearpage}
\begin{appendix}[A]
\section*{\begin{center}Derivation of the PDF of $Y$\end{center}}

%
%\begin{equation} 
%\label{appA1}
%\begin{array}{ll}
%\left[\,Y\mid f\,\right] =\mathcal{N}(\mu,\Sig)\\
%\left[\,Y\mid \overline{f}\,\right] =\mathcal{N}(\overline{\mu},\Sig)\\
%\phi^*(Y) = (\mu-\overline{\mu})'\Sig^{-1}Y\\
%u^*=\frac{1}{2}(\mu-\overline{\mu})'\Sig^{-1}(\mu+\overline{\mu})\\
%Y = \x\beta + \varepsilon\\
%\Var(\varepsilon) = \mathbf C + \mathbf R\\
%Y = \mu + \eta\\
%\Var(\eta) = \mathbf C + \mathbf R + \mathbf Q\\
%\mathbf Q = \x \Var(\beta)\x' =\sigma^2\x\x'\\
%\mu = \x\Esp(\beta)=\sum_{i=1}^p x_i\\
%\phi^*(Y) = \phi^{m}(Y) + \phi^{a}(Y) 
%%&=\mathcal{N}(\x e,\mathbf C+\mathbf R+\x\Om\x')
%\end{array}
%\end{equation}
%
%

To obtain Equation (\ref{gauss3}) from Equation (\ref{gauss1}) and (\ref{gauss2}), we integrate out $\beta$:
\begin{equation} 
\label{appA1}
\begin{array}{ll}
\left[\,Y\mid \x,\mathbf{C}, \mathbf{R}\,\right] =  \int_\beta\,\left[\,Y\mid \beta,\x,\mathbf{C}, \mathbf{R}\,\right]\,.\,\left[\,\beta\mid\omega\,\right]\,\textrm{d}\beta\\
%&=\mathcal{N}(\x e,\mathbf C+\mathbf R+\x\Om\x')
\end{array}
\end{equation}
Given the quadratic dependence to $\beta$ of the two terms under the integral in the right hand side of Equation (\ref{appA1}), it is clear that the PDF of the left hand side is also Gaussian. Thus, instead of computing the above integral, it is more convenient to derive the mean and variance of this PDF by applying the rule of total expectation and total variance:
\begin{equation} 
\label{appA2}
\begin{array}{ll}
\Esp(Y\mid \x,\mathbf{C}, \mathbf{R})&=\Esp\left(\Esp(Y\mid \beta, \x,\mathbf{C}, \mathbf{R})\mid \x,\mathbf{C}, \mathbf{R}\right)=\Esp\left(\x\beta\mid \x,\mathbf{C}, \mathbf{R}\right)=\x\Esp\left(\beta\right)\\
&=\x e\\
\Var(Y\mid \x,\mathbf{C}, \mathbf{R})&=\Var\left(\Esp(Y\mid \beta, \x,\mathbf{C}, \mathbf{R})\mid \x,\mathbf{C}, \mathbf{R}\right)+\Esp\left(\Var(Y\mid \beta, \x,\mathbf{C}, \mathbf{R})\mid \x,\mathbf{C}, \mathbf{R}\right)\\
&=\Var\left(\x\beta\mid \x,\mathbf{C}, \mathbf{R}\right)+\Esp\left(\mathbf C+\mathbf R\mid \x,\mathbf{C}, \mathbf{R}\right)\\
&=\x\Var(\beta)\x'+\mathbf C+\mathbf R=\omega^2\x\x'+\mathbf C+\mathbf R\\
\left[\,Y\mid \x,\mathbf{C}, \mathbf{R}\,\right]& =\mathcal{N}(\x e,\mathbf{C}+\mathbf{R}+\omega^2\x\x')\\
\end{array}
\end{equation}

Next, in order to account for the sampling uncertainty on the estimation of $\mu$, we apply Bayes theorem to derive the PDF of $\mu$ conditional on the ensemble $\E$. Denote $\mu^{(1)},...,\mu^{(r)}$ the $r$  simulated responses in $\E$ which are assumed to be i.i.d. according to a Gaussian with mean $\mu$ and covariance $\mathbf C$. We have:
\begin{equation} 
\label{modelcov1}
\begin{array}{ll}
\left[\,\mu\mid \mathbf{C}, \E\,\right]&\propto \Pi_{j=1}^r \left[\,\mu^{(j)}\mid \mathbf{C}\,\right]\,.\,\left[\,\mu\,\right]\\
&\propto \Pi_{j=1}^r \,\,\mathcal{N}(\mu^{(j)}\mid \mu,\mathbf{C})\\
&=\mathcal{N}(\mu\mid \widehat{\mu},\frac{1}{r}\mathbf{C})
\end{array}
\end{equation}
where $\widehat{\mu}$ is the empirical mean of the ensemble, and we use the improper prior $\left[\,\mu\,\right]\propto 1$. The exact same approach yields $\left[\,x_i\mid \mathbf{C}, \E\,\right]\propto \Pi_{j=1}^{r_i} \,\,\mathcal{N}(\x_i^{(j)}\mid x_i,\mathbf{C})=\mathcal{N}(x_i\mid \widehat{x}_i,\frac{1}{r_i}\mathbf{C})$.

To integrate out $\mu$, we proceed by following the same reasoning as above for integrating out $\beta$. Since the resulting PDF is clearly Gaussian, it suffices to derive its mean and variance:
\begin{equation} 
\label{appA5}
\begin{array}{ll}
\Esp(Y\mid \x,\mathbf{C}, \mathbf{R},\E)&=\Esp\left(\Esp(Y\mid \mu, \x,\mathbf{C}, \mathbf{R},\E)\mid \x,\mathbf{C}, \mathbf{R},\E\right)=\Esp\left(\mu\mid \x,\mathbf{C}, \mathbf{R},\E\right)\\
&=\widehat{\mu}\\
\Var(Y\mid \x,\mathbf{C}, \mathbf{R},\E)&=\Var\left(\Esp(Y\mid \mu, \x,\mathbf{C}, \mathbf{R},\E)\mid \x,\mathbf{C}, \mathbf{R},\E\right)+\Esp\left(\Var(Y\mid \mu, \x,\mathbf{C}, \mathbf{R},\E)\mid \x,\mathbf{C}, \mathbf{R},\E\right)\\
&=\Var\left(\mu\mid \x,\mathbf{C}, \mathbf{R}\right)+\Esp\left(\omega^2\x\x'+\mathbf C+\mathbf R\mid \x,\mathbf{C}, \mathbf{R},\E\right)\\
&=\frac{1}{r}\mathbf C+\omega^2\x\x'+\mathbf C+\mathbf R\\
\end{array}
\end{equation}
Likewise, to integrate out $\x$, we derive the total mean and total variance:
\begin{equation} 
\label{modelcov1}
\begin{array}{ll}
\Esp(Y\mid \mathbf{C}, \mathbf{R},\E)&=\Esp\left(\Esp(Y\mid \x,\mathbf{C}, \mathbf{R},\E)\mid \mathbf{C}, \mathbf{R},\E\right)=\Esp\left(\widehat{\mu}\mid \mathbf{C}, \mathbf{R},\E\right)\\
&=\widehat{\mu}\\
\Var(Y\mid \mathbf{C}, \mathbf{R},\E)&=\Var\left(\Esp(Y\mid \x, \mathbf{C}, \mathbf{R},\E)\mid \mathbf{C}, \mathbf{R},\E\right)+\Esp\left(\Var(Y\mid \x, \mathbf{C}, \mathbf{R},\E)\mid \mathbf{C}, \mathbf{R},\E\right)\\
&=\mathbf 0 +(1+\frac{1}{r})\mathbf{C}+\mathbf R+\Esp\left( \omega^2\x\x'\mid \mathbf{C},\E\right)\\
&=(1+\frac{1}{r})\mathbf{C}+\mathbf R+\omega^2\sum_{i}\Esp\left( x_i\,x_i'\mid \mathbf{C},\E\right)\\
&=(1+\frac{1}{r})\mathbf{C}+\mathbf R+\omega^2\sum_{i}\Var\left( x_i\mid \mathbf{C},\E\right)+\omega^2\sum_{i}\Esp\left( x_i\mid \mathbf{C},\E\right)\Esp\left( x_i\mid \mathbf{C},\E\right)'\\
&=(1+\frac{1}{r})\mathbf{C}+\mathbf R+\omega^2\sum_{i}\frac{1}{r_i}\mathbf C+\omega^2\sum_{i}\widehat{x}_i\,\widehat{x}_j'\\
&=(1+\frac{1}{r}+\omega^2\sum_i\frac{1}{r_i})\mathbf{C}+\mathbf R+\omega^2\widehat{\x}\widehat{\x}'\\
&=\mathbf{C}+\mathbf R+\omega^2\widehat{\x}\widehat{\x}'+\lambda \mathbf{C}\\
\end{array}
\end{equation}
with $\lambda=1/r+\omega^2\sum_i1/r_i$. Note that $\left[\,Y\mid \mathbf{C}, \mathbf{R},\E\,\right]$ is no longer Gaussian after integrating out $\x$, because $\x$ appears in the covariance of $\left[\,Y\mid \x, \mathbf{C}, \mathbf{R},\E\,\right]$. However, for simplicity, we approximate it to be Gaussian. 

The sampling uncertainty on the covariance matrix $\mathbf C$ is addressed by using an approach described in \cite{HN14} which main ideas are succinctly recalled here. The reader is referred to the publication for details and explicit calculations. In summary, we apply Bayes theorem in order to derive $\left[\,\mathbf C\mid \mathbf \E \,\right]$, as for $\mu$ and $\x$. However, we use this time an informative conjugate prior on $\mathbf C$, as opposed to an improper prior.
\begin{equation} 
\label{modelcov1}
\begin{array}{ll}
\left[\,\mathbf C\mid \mathbf \Delta, a\,\right] = \mathcal{IW}(\mathbf \Delta, a)
\end{array}
\end{equation}
where $\mathbf \Delta$ denotes the a priori mean of $\mathbf C$ and $a$ is a scalar parameter that drives the a priori variance.  Furthermore, the mean and variance parameters $(\mathbf \Delta,a)$ of this informative prior are estimated from $\E$ by maximizing the marginal likelihood $\ell(a,\D)$ associated to this Bayesian model. 
\begin{linenomath*}\begin{equation}
\label{ell}
\begin{array}{lll}
\ell(a,\D)&=\,\,\,(\frac{a\,r_0}{1-a}+n+1)\log|\frac{a}{1-a}\mathbf\Delta|-(\frac{r_0}{1-a}+n+1)\log|\Sn+\frac{a}{1-a}\mathbf{\Delta}|\\%+K(a)\\
%\textrm{with    }K(a)=
&+\,\,\,\,2\log\left(\Gamma_n\{\frac{1}{2}(\frac{r_0}{1-a}+n+1)\}\,/\,\Gamma_n\{\frac{1}{2}(\frac{a\,r_0}{1-a}+n+1)\}\right).
\end{array}
\end{equation}\end{linenomath*}
where $\Gamma_n$ is the $n-$variate Gamma function and $\widehat{\Om}=\x_0\x_0'/r_0$ is the empirical covariance. The estimators $(\widehat{a},\widehat{\mathbf \Delta})$ satisfy to:
\begin{linenomath*}\begin{equation}
\label{maxll}
\begin{array}{ll}
(\widehat{a},\widehat{\D})=\textrm{argmax}_{a\in[0,1],\,\, \D\in\mathcal{F}}\,\, \ell(a,\D), \,\,
\end{array}
\end{equation}\end{linenomath*}
where $\mathcal{F}$ is a set of definite positive matrices chosen to introduce a regularization constraint on the covariance. Here we choose $\mathcal{F}=\{\textrm{diag}(\delta_1,...,\delta_n)\mid \delta_1>0,...,\delta_n>0\}$ the set of definite positive diagonal matrices, and we derive an approximated solution to Equation (\ref{maxll}) with $\widehat{\D}=\textrm{diag}(\widehat{\Om})$ and $\widehat{a}=\textrm{argmax}_{a\in[0,1]}\,\, \ell(a,\widehat\D)$. Because the prior PDF is fitted on the data, this approach can be referred to as ``empirical bayesian''.  The ``fitted'' prior $[\,\mathbf C\mid \widehat{\mathbf{\Delta}}, \widehat a\,]$ is then updated using the ensemble $\E$, and the obtained posterior has a closed form expression due to conjugacy:
\begin{equation} 
\label{modelcov1}
\begin{array}{ll}
\left[\,\mathbf C\mid \mathbf \E, \widehat{\mathbf{\Delta}}, \widehat a\,\right] &\propto \left[\,\E\mid \mathbf C\,\right]\,.\,\mathcal{IW}(\widehat{\mathbf{\Delta}}, \widehat a)=\mathcal{IW}( \widehat{\mathbf C}, \widehat{a}')
\end{array}
\end{equation}
where $\widehat{\mathbf C}=\widehat a \widehat{\mathbf\Delta}+(1-\widehat a)\widehat{\Om}$ and $\widehat{a}'=1/(2-\widehat{a})$.
We can then use the above posterior to integrate out $\mathbf C$ in the PDF of $Y$, in order to obtain $[\,Y\mid \E, \mathbf R, \widehat{\mathbf{\Delta}}, \widehat a\,]$:
\begin{equation} 
\label{appstud}
%\begin{array}{ll}
\left[\,Y\mid \E,\mathbf R, \widehat{\mathbf{\Delta}}, \widehat a\,\right] =\int_{\mathbf C}\,\left[\,Y\mid \mathbf C,\mathbf R,\E\,\right]\,.\,\left[\,\mathbf C\mid \E,\widehat{\mathbf{\Delta}}, \widehat a\,\right]\,\textrm{d}\mathbf C\\
%&= \mathcal{S}t(\widehat{\mu},\sigma^2 \mathbf I+\omega^2\widehat{\x}\widehat{\x}'+(1+\lambda)\widehat{\mathbf{C}},\widehat{\nu})
%\end{array}
\end{equation}
The integral above does not have a closed form expression because the variance $\Sig=\mathbf R+\omega^2\widehat{\x}\widehat{\x}'+(1+\lambda)\mathbf{C}$ of $[\,Y\mid \mathbf C,\mathbf R,\E\,]$ is not proportional to $\mathbf C$.  To address this issue, we approximate $[\,\mathbf \Sig\mid \mathbf \E, \widehat{\mathbf{\Delta}}, \widehat a\,]$ by $\mathcal{IW}( \mathbf R+\omega^2\widehat{\x}\widehat{\x}'+(1+\lambda)\widehat{\mathbf{C}}, \widehat{a}')$. This assumption is conservative in the sense that it extends the sampling uncertainty on $\mathbf C$ to $\mathbf R+\omega^2\widehat{\x}\widehat{\x}'+(1+\lambda)\mathbf{C}$ even though $\mathbf R+\omega^2\widehat{\x}\widehat{\x}'$ is a constant. It yields a closed form expression of the above integral thanks to conjugacy:
\begin{equation} 
\label{appstud2}
%\begin{array}{ll}
\left[\,Y\mid \E,\mathbf R, \widehat{\mathbf{\Delta}}, \widehat a\,\right] = \mathcal{S}t(\widehat{\mu},\mathbf R+\omega^2\widehat{\x}\widehat{\x}'+(1+\lambda)\widehat{\mathbf{C}},\widehat{\nu})
%\end{array}
\end{equation}

\end{appendix}

\ifthenelse{\boolean{dc}}
{}
{\clearpage}
\begin{appendix}[B]
\section*{\begin{center}Optimal index derivation\end{center}}
Let us solve the optimization problem of Equation (\ref{event2}) under the above assumptions. 
For simplicity, we restrict our search to so called ``\textit{half-space}'' events which are defined by $E=\{Y\in\Omega_f\mid \phi'Y\geq u\}$ where $\phi'Y$ is a linear index with $\phi$ a vector of dimension $n$, and $u$ is a threshold. Let us consider PNS as a function of $\phi$ and $u$.% Since $\phi'Y$ follows a Gaussian distribution with mean $A'\mu$ in the factual world and $A'\overline{\mu}_f$ in the counterfactual world and with variance $A'\Sig A$ in both worlds, we have for any given $A$:
\begin{equation}
\begin{array}{ll}
\PNS(\phi,u)&=\Proba(\phi'Y\geq u\mid f)-\Proba(\phi'Y\geq u\mid \overline{f})\\
%&=\Phi\left(\frac{u-A'\overline{\mu}_f}{\sqrt{A'\Sig A}}\right)-\Phi\left(\frac{u-A'\mu}{\sqrt{A'\Sig A}}\right)
\end{array}
\end{equation}
For simplicity, we will use an expression of $\PNS(\phi,u)$ in the treatment of the optimization problem which approximates $[\phi'Y\mid f]$ by a Gaussian PDF, even though it is a Student $t$ PDF from the calculations of Section 4. Note that this approximation is made restrictively here for deriving an optimal index. Once this index is obtained, it is the then the true Student $t$ PDF of $Y$ that will be used to derive the desired value of $\PNS$. Therefore, the implication of this approximation is to yield an index which is suboptimal and thereby underestimates the maximized value $\PNS^*$. 
\begin{equation}
\begin{array}{ll}
\PNS(\phi,u)&=F\left(\frac{u-\phi'\overline{\mu}}{\sqrt{\phi'\overline{\Sig} \phi}}\right)-F\left(\frac{u-\phi'\mu}{\sqrt{\phi'\Sig \phi}}\right)
\end{array}
\end{equation}
where $F$ is the standard Gaussian CDF. The first order condition in $u$, $\partial \PNS(\phi,u)/\partial u =0$, thus yields:
\begin{equation}
\begin{array}{ll}
\label{B3}
&\exp\left(-\frac{(u-\phi'\overline{\mu})^2}{2\phi'\overline{\Sig} \phi}\right)=\exp\left(-\frac{(u-\phi'\mu)^2}{2\phi'\Sig \phi}\right)\\
\end{array}
\end{equation}
Next, we introduce a third approximation $\Sig\simeq\overline{\Sig}$ to solve Equation (\ref{B3}),  yielding:
\begin{equation}
\begin{array}{ll}
u^*=\frac{1}{2}\phi'(\mu+\overline{\mu})\\
\Rightarrow \PNS(\phi,u^*)=2F\left(\frac{\phi'(\mu-\overline{\mu})}{2\sqrt{\phi'\Sig \phi}}\right)-1
\end{array}
\end{equation}
Then, the first order condition in $\phi$, $\partial \PNS(\phi,u^*)/\partial \phi =0$,  yields:
\begin{equation}
\begin{array}{ll}
&(\phi'\Sig \phi)(\mu-\overline{\mu})=(\phi'(\mu-\overline{\mu}))\Sig \phi\\
\Rightarrow&\phi^*=\Sig^{-1}(\mu-\overline{\mu})\\
\end{array}
\end{equation}
which proves Equation (\ref{optphi}). Figure 5c illustrates this solution and also shows that the optimization problem of Equation (\ref{event2}) may be viewed as a classification problem. Our proposal to solve Equation (\ref{event2}) is in fact  similar to a commonplace classification algorithm used in machine learning and known as Support Vector Machine (SVM) \citep{CoV95}.
\end{appendix}

\ifthenelse{\boolean{dc}}
{}
{\clearpage}
\begin{appendix}[C]
\section*{\begin{center}Data used in illustration\end{center}}
As in \cite{Han16}, observations were obtained from the HADCRUT4 monthly temperature dataset \citep{Mo12}, while GCM model simulations were obtained from the IPSL CM5A-LR model \citep{Duf12}, downloaded from the CMIP5 database. An ensemble of runs consisting of two sets of forcings was used, the natural set of forcings (NAT) and the anthropogenic set of forcings (ANT) for which three runs are available in each case over the period of interest and from which an ensemble average was derived. On the other hand, a single preindustrial control run of 1000 years is available and was thus split into ten individual control runs of 100 years. Temperature in both observations and simulations were converted to anomalies by subtracting the time average over the reference period 1960-1991. The data was averaged temporally and spatially using a temporal resolution of ten years. Averaging was performed for both observations and simulations by using restrictively values for which observations were non missing, for a like-to-like comparison between observations and simulations.
\end{appendix}

% Create a bibliography directory and place your .bib file there.
% -REMOVE ALL DIRECTORY PATHS TO REFERENCE FILES BEFORE SUBMITTING TO THE AMS FOR PEER REVIEW
\ifthenelse{\boolean{dc}}
{}
{\clearpage}
\bibliographystyle{ametsoc}
\bibliography{references}

%%%%%%%%%%%%%%%%%%%%%%%%%%%%%%%%%%%%%%%%%%%%%%%%%%%%%%%%%%%%%%%%%%%%%
% FIGURES-REMOVE ALL DIRECTORY PATHS TO FIGURE FILES BEFORE SUBMITTING TO THE AMS FOR PEER REVIEW
%%%%%%%%%%%%%%%%%%%%%%%%%%%%%%%%%%%%%%%%%%%%%%%%%%%%%%%%%%%%%%%%%%%%%

\begin{figure}[t]
\begin{center} 
\noindent\includegraphics[angle=0, width=5in]{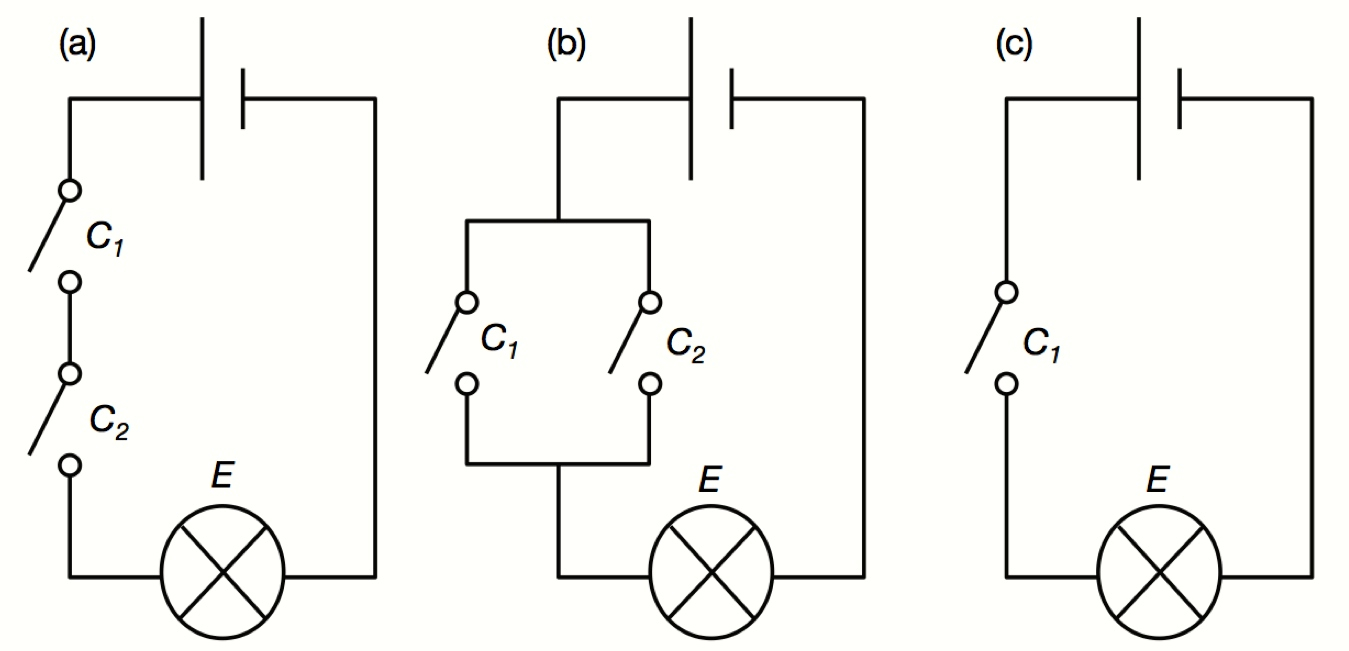}
\end{center} 
\caption{The three facets of causality. (a) Bulb $E$ can never be lit unless switch $C_1$ is on, yet activating $C_1$ does not always result in lighting $E$ as this also requires turning on $C_2$: turning on $C_1$ is thus a necessary cause of $E$ lighting, but not a sufficient one. (b) $E$ is lit any time $C_1$ is turned on, yet if $C_1$ is turned off $E$ may still be lit by activating $C_2$: turning on $C_1$ is thus a sufficient cause of $E$ lighting, but not a necessary one. (c) Turning on $C_1$ always lights $E$, and $E$ may not be lighted unless $C_1$ is on: turning on $C_1$ is thus a necessary and sufficient cause of $E$ lighting.}
\label{f1}
\end{figure}

%\begin{figure}[t]
%\noindent\includegraphics[angle=0, width=6in]{multiv_fingerprint.pdf}\\
%\caption{Derivation of PNS for $n=2$. (a) Scatterplots of $(Y_1,Y_2)$ for the factual (red dots) and counterfactual (green dots) ensembles; optimal fingerprint event $E_y$ of Eq. (\ref{event}) obtained under the linear Gaussian setting; observed value $(y_1,y_2)$ (black diamond). (b) Factual PDF (green line) and counterfactual PDF (red line) of the optimal fingerprint index $Z$; observed value. (c) PNS as a function of the threshold on $Z$.}
%\label{f2}
%\end{figure}

\begin{figure}[t]
\begin{center} 
\noindent\includegraphics[angle=0, width=6.6in]{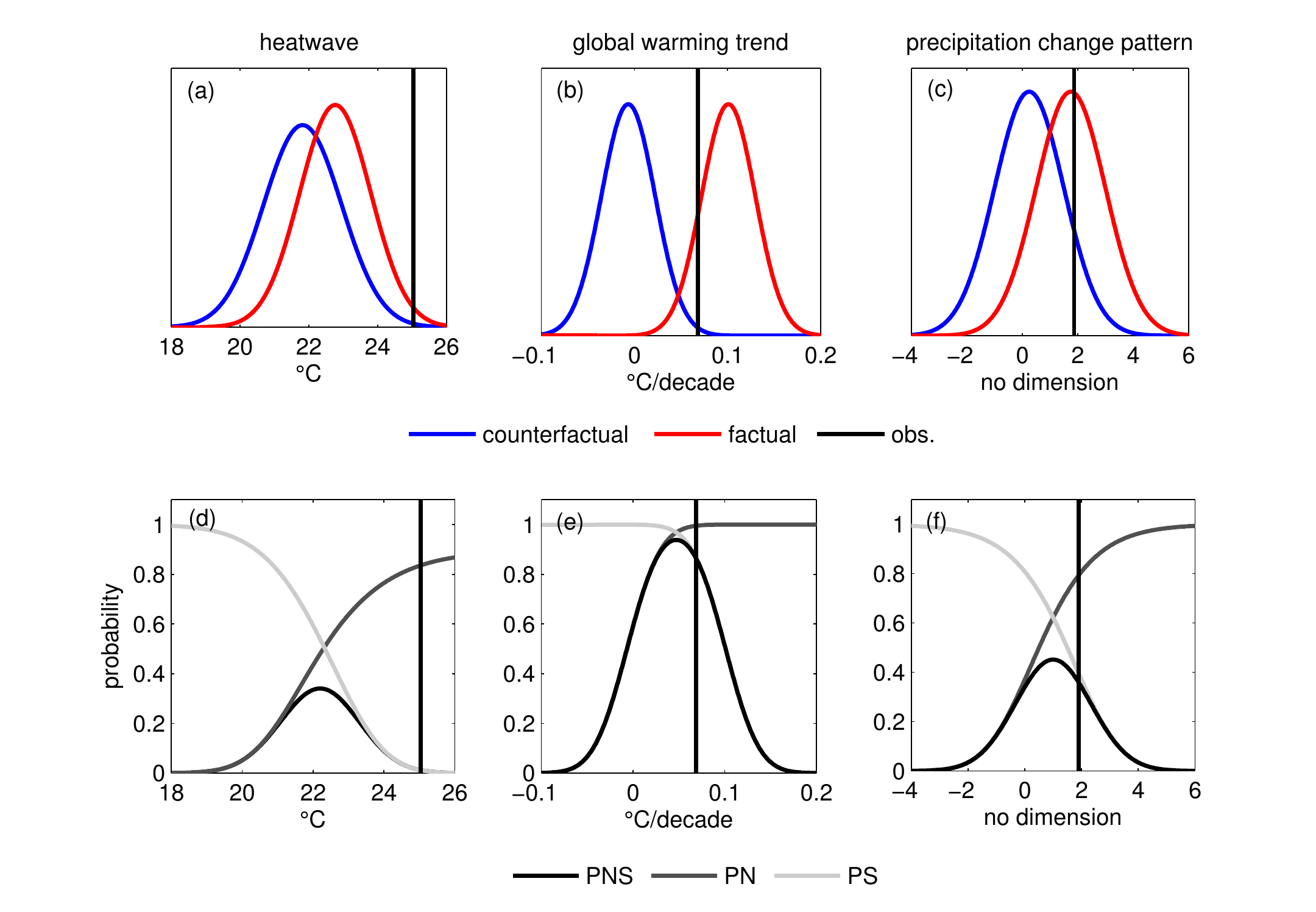}\\
\end{center} 
\caption{Probabilities of causation in three different climate attribution situations. Upper panels (a,b,c) : factual PDF (red line) and counterfactual PDF (blue line) of the relevant index $Z$, observed value $z$ of the index (vertical black line). Lower panels (d,e,f): $\PN$, $\PS$ and $\PNS$ for the event $\{Z\geq u\}$ as a function of the threshold $u$. Left column (a,d): attribution of the Argentinian heatwave of December 2013. Middle column (b,e): attribution of the 20th century temperature change. Left column (c,f): attribution of the precipitation change over the satellite era \citep{MB13}.}
\label{f2b}
\end{figure}

\begin{figure}[t]
\begin{center} 
\noindent\includegraphics[angle=0, width=3.5in]{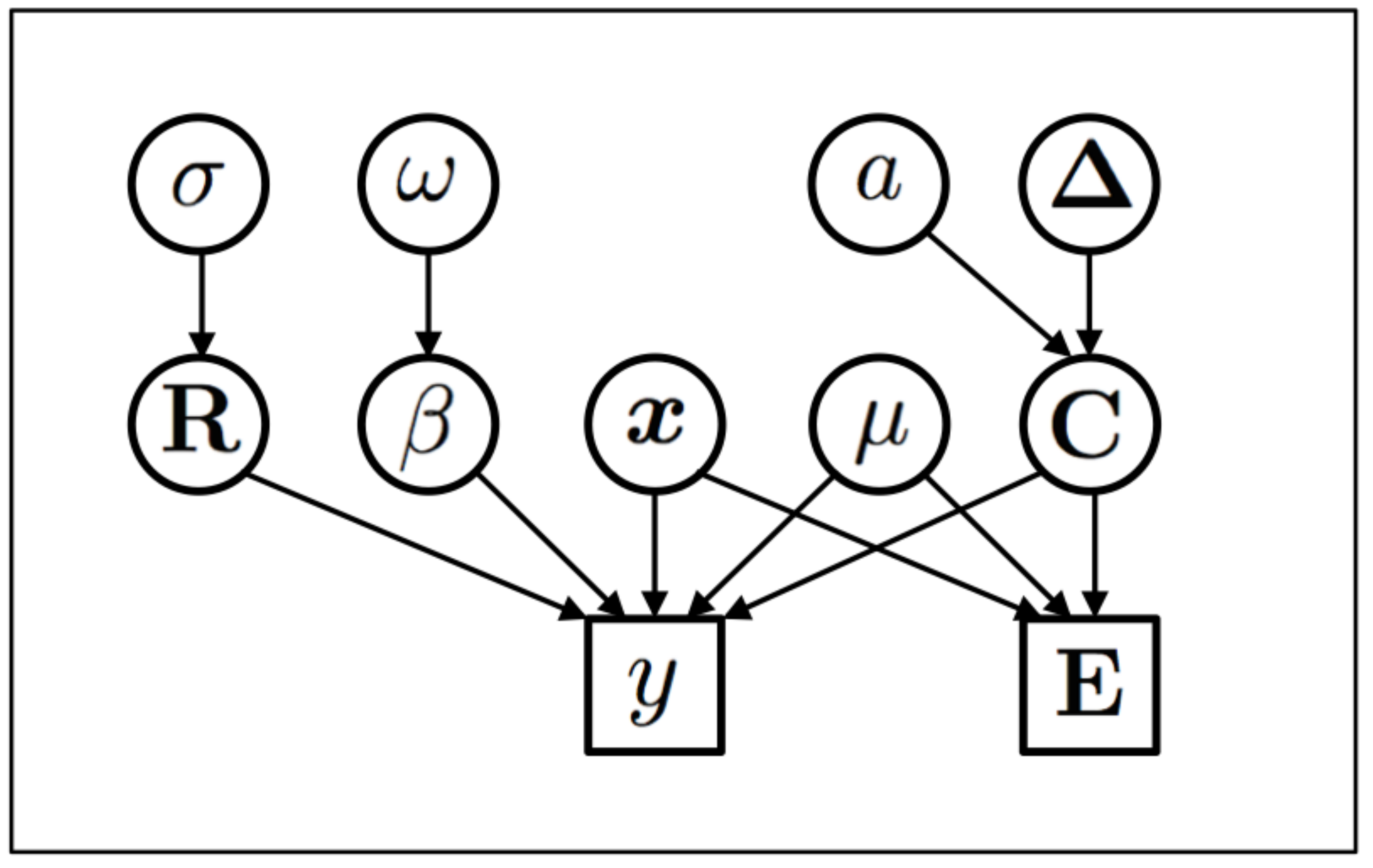}\\
\end{center} 
\caption{Structural chart of the statistical model introduced in Section 4: underlying hierarchy of parameters (i.e. unobserved quantities, circles); and data used for inference (i.e. observed quantities, squares).}
\label{f2b}
\end{figure}

%\begin{figure}[t]
%\noindent\includegraphics[angle=0, width=5.5in]{phi.pdf}\\
%\caption{PNS as a function of the signal to noise ratio under the linear Gaussian setting. (a) Linear scale (vertical axis). (b) Logarithmic scale (vertical axis).}
%\label{f3}
%\end{figure}

\begin{figure}[t]
\begin{center} 
\noindent\includegraphics[angle=0, width=3.2in]{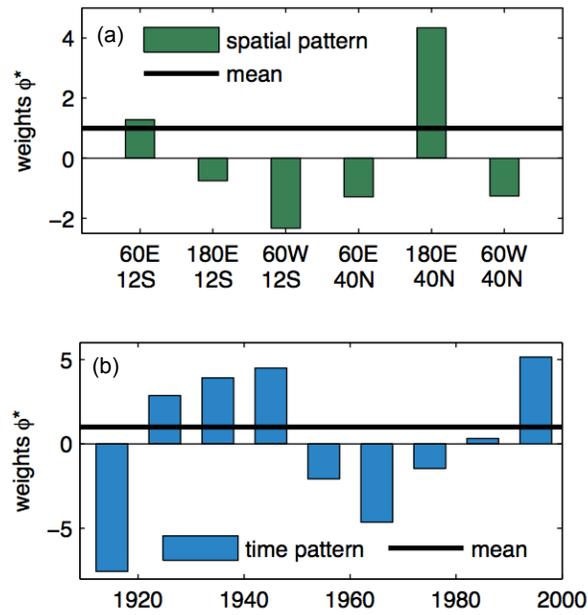}\\
\end{center} 
\caption{Illustration on the 20th century temperature change: optimal mapping $\phi^*$. (a) components of $\phi^*$ averaged spatially. (b) components of $\phi^*$ averaged temporally. }
\label{f4}
\end{figure}

\begin{figure}[t]
\begin{center} 
\noindent\includegraphics[angle=0, width=5in]{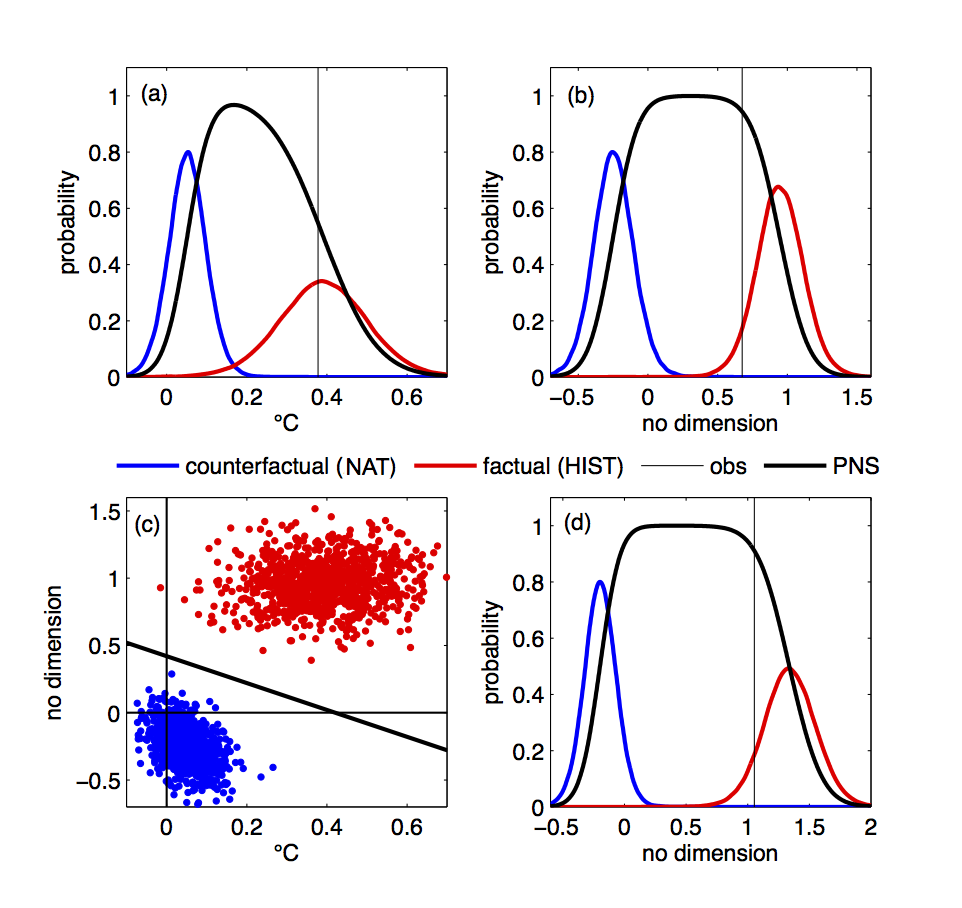}\\
\end{center} 
\caption{Illustration on the 20th century temperature change: results. (a) Factual PDF (red line) and counterfactual PDF (blue line) of the global mean index, observed value (thin vertical black line); $\PNS$ as a function of the threshold $u$ (thick black line). (b) Same as (a) for the space-time pattern index. (c) Scatterplot of factual (red dots) and counterfactual (blue dots) joint realizations of the global mean index (horizontal axis) and of the space-time pattern index (vertical axis). (d) Same as (a) for the optimal index $Z=\phi^*(Y)$.}
\label{f4}
\end{figure}

\begin{figure}[t]
\begin{center} 
\noindent\includegraphics[angle=0, width=3.8in]{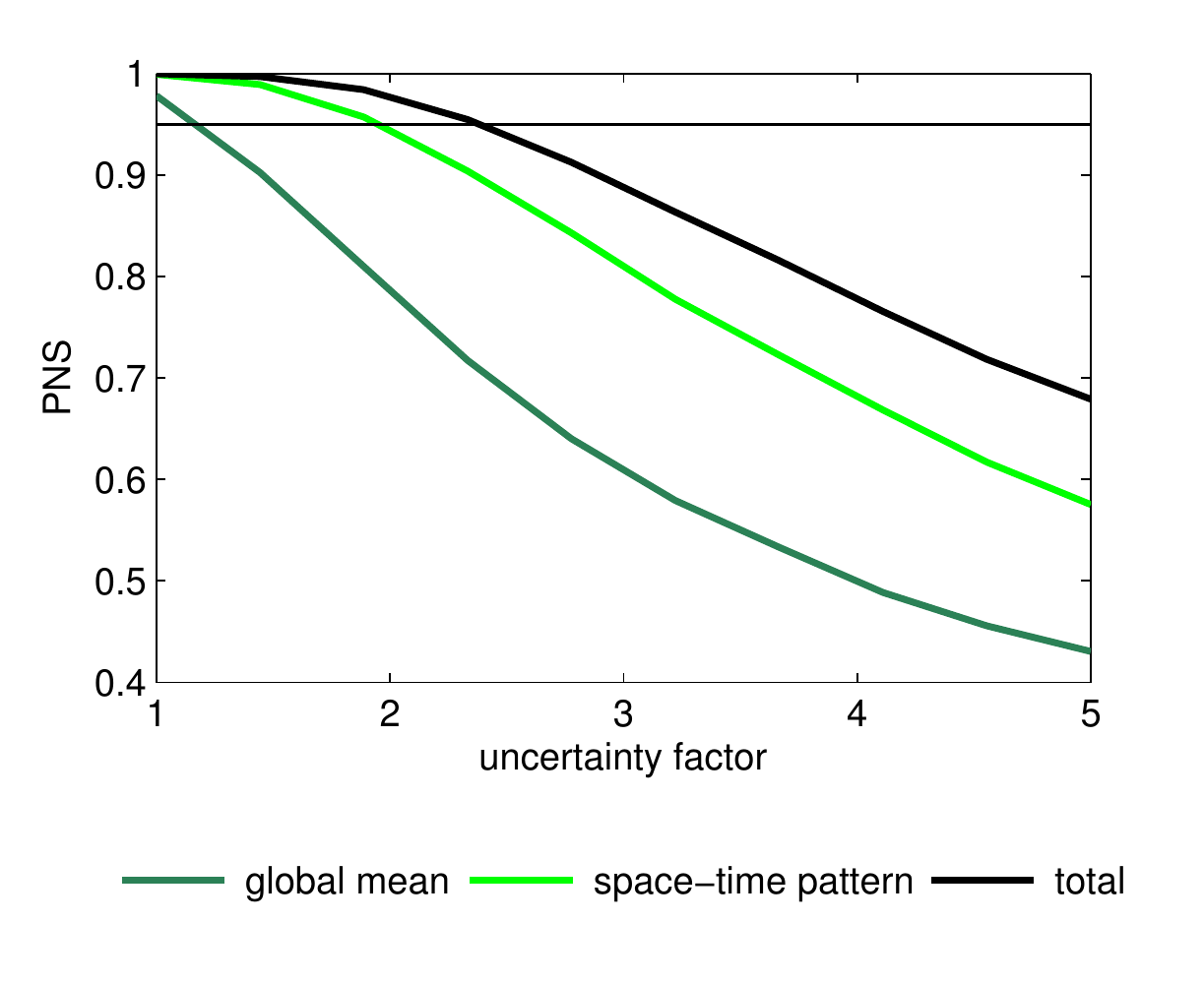}\\
\end{center} 
\caption{$\PNS$ as a function of the inflation factor applied to all uncertainty sources: global mean alone (light green line), space-time pattern (dark green line), total (thick black line).}
\label{f4}
\end{figure}

\begin{figure}[t]
\begin{center} 
\noindent\includegraphics[angle=0, width=5in]{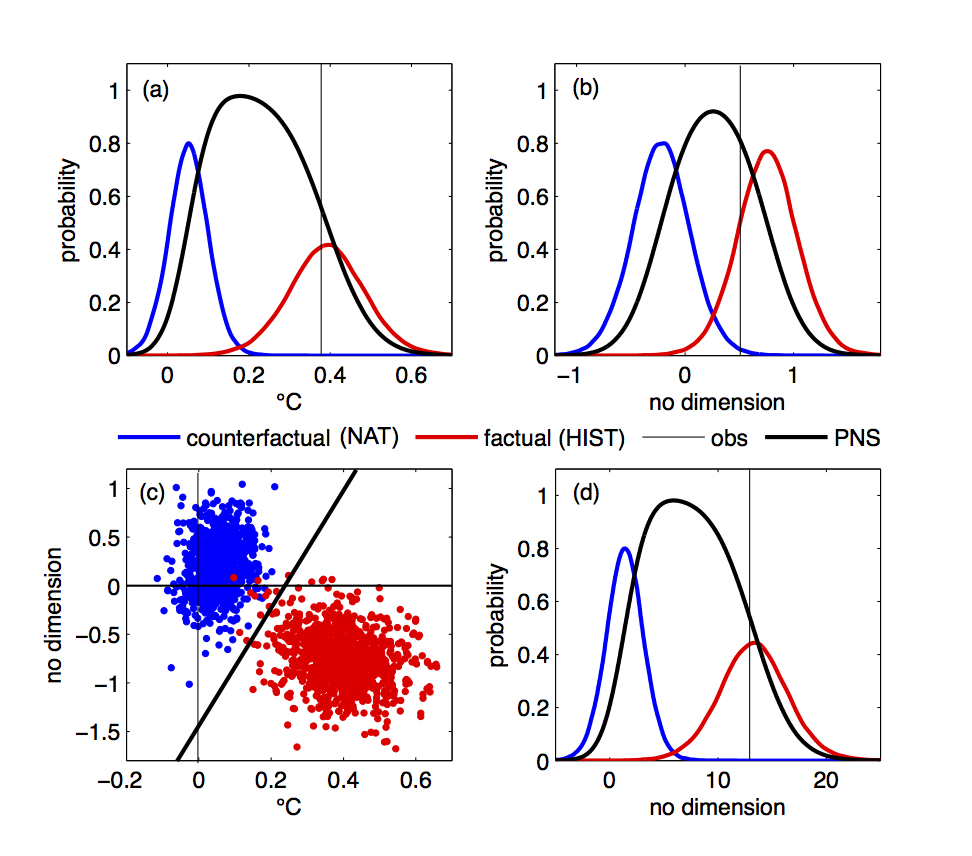}\\
\end{center} 
\caption{Same as Figure 5 for the mapping $\phi^+$ projected onto the leading eigenvectors of $\mathbf C$.}
\label{f4}
\end{figure}

\begin{figure}[t]
\begin{center} 
\noindent\includegraphics[angle=0, width=3.2in]{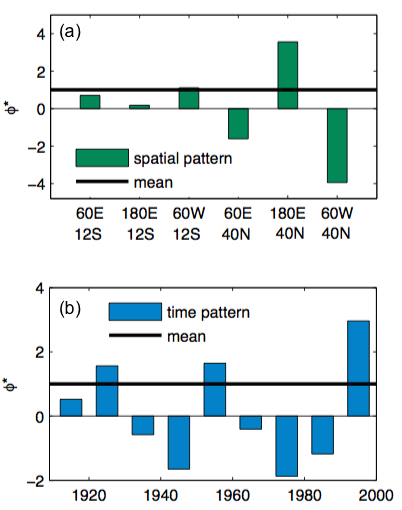}\\
\end{center} 
\caption{Same as Figure 4 for the mapping $\phi^+$ projected onto the leading eigenvectors of $\mathbf C$.}
\label{f4}
\end{figure}

%%%%%%%%%%%%%%%%%%%%%%%%%%%%%%%%%%%%%%%%%%%%%%%%%%%%%%%%%%%%%%%%%%%%%
% TABLES
%%%%%%%%%%%%%%%%%%%%%%%%%%%%%%%%%%%%%%%%%%%%%%%%%%%%%%%%%%%%%%%%%%%%%

\begin{table}[t]
\label{t1}
\begin{center}
\begin{tabular}{|l|l|}
\hline
\textbf{Term} & \textbf{Probability}\\
\hline
\textit{Virtually certain} & $\geq0.99$\\
\textit{Extremely likely} & $\geq0.95$\\
\textit{Very likely} & $\geq0.90$\\
\textit{Likely} & $\geq0.66$\\
\textit{About as likely as not} & $>0.33$ and $<0.66$\\
\textit{Unlikely} & $\leq0.33$\\
\textit{Very unlikely} &$\leq0.10$ \\
\textit{Exceptionally unlikely} & $\leq0.01$\\
\hline
\end{tabular}
\end{center}
\caption{Correspondence between language and probabilities in IPCC calibrated terminology \citep{Mas10}.}
\end{table}
\end{document}